\documentclass[twocolumn,showpacs,preprintnumbers,amsmath,amssymb]{revtex4}
\usepackage{amsmath,amssymb,color,epsfig,latexsym,natbib,graphicx,dcolumn}
\usepackage{bm}
\bibpunct{[}{]}{,}{n}{,}{,}             
  
\def\DEL#1{{\textcolor{green}{}}}         

\def\rms{r.m.s.}

\newcommand{\bB}{\bf{B}}

\newcommand\Rv{\ensuremath{\sf R\mspace{-2mu}v}}

\newcommand{\Rla}{R_{\lambda}}
\newcommand{\rem}[1]{}

\newcommand\vecp[1]{\vec{#1}}                   

\newcommand{\be}{\begin{equation}}
\newcommand{\ee}{\end{equation}}

\def\bB0{\vecp{B}_0}   

\begin{document}
\title{\bf Spectral Modeling of Turbulent Flows and the Role of Helicity } 
\author{J. Baerenzung}
\affiliation{Laboratoire Cassiop\'ee, UMR 6202, Observatoire de la C\^ote d'Azur, B.P. 4229, 06304 Nice Cedex 4, France }
\author{H. Politano}
\affiliation{Laboratoire Cassiop\'ee, UMR 6202, Observatoire de la C\^ote d'Azur, B.P. 4229, 06304 Nice Cedex 4, France }
\author{Y. Ponty}
\affiliation{Laboratoire Cassiop\'ee, UMR 6202, Observatoire de la C\^ote d'Azur, B.P. 4229, 06304 Nice Cedex 4, France }
\author{A. Pouquet}
\affiliation{TNT/NCAR, P.O. Box 3000, Boulder, Colorado 80307-3000, U.S.A.}

\begin{abstract}
We present a new version of a dynamical spectral model for Large Eddy Simulation based on the 
Eddy Damped Quasi Normal Markovian approximation \cite{sao,chollet_lesieur}. 
Three distinct modifications are implemented and tested. On the one hand, whereas in current approaches, 
a Kolmogorov-like energy spectrum is usually assumed in order
to evaluate the nonlocal transfer, in our method the energy spectrum of the subgrid scales adapts itself dynamically to the 
large-scale resolved spectrum; this first modification allows in particular for a better treatment of 
transient phases and instabilities, as shown on one specific example. 
Moreover, the model takes into account the phase relationships of the small-scales, 
embodied for example in strong localized structures such as vortex filaments. 
To that effect, phase information is implemented in the treatment of the so-called eddy noise in the closure model.
Finally, we also consider the role that helical small scales may play  in the evaluation of the transfer
of energy and helicity, the two invariants of the primitive equations in the inviscid case; this leads
as well to intrinsic variations in the development of helicity spectra. 
Therefore, our model allows for simulations of flows for a variety of circumstances and a priori 
at any given Reynolds number. 
Comparisons with Direct Numerical Simulations of the three-dimensional Navier-Stokes equation are performed 
on fluids driven by an ABC (Beltrami) flow which is a prototype of fully helical flows. 
Good agreements are obtained for physical and spectral behavior of the large scales. 
\end{abstract}
\maketitle

\section{Introduction}
Turbulent flows are ubiquitous, and they are linked to many issues in the geosciences, as in meteorology, 
oceanography, climatology, ecology, solar--terrestrial interactions and fusion, as well as the generation 
and ensuing dynamics of magnetic fields in planets, 
stars and galaxies due to e.g. convective fluid motions. As manifestations of one of the last 
outstanding unsolved problems of classical physics, 
such flows form today the focus of numerous investigations. 

Natural flows are often in a turbulent state driven by large scale forcing (novae explosions in the interstellar medium) 
or by instabilities (convection in the sun). Such flows involve a huge number of coupled modes at different scales leading to 
great complexity both in their temporal dynamics and in their emerging physical structures.
Nonlinearities prevail when the Reynolds number $\Rv$ -- which measures the amount of active temporal or 
spatial scales in the problem -- is large. 
In the Kolmogorov framework  \cite{K41}, the number of degrees of freedom increases as $\Rv^{9/4}$ for $\Rv\gg1$;
for example in geophysical flows, $\Rv$ is often larger than $10^8$. 
The ability to probe large $\Rv$, and to examine in details the 
large-scale behavior of turbulent flows depends critically on the ability to resolve such a large number of spatial and 
temporal scales, or else to model them adequately.

Only modest Reynolds numbers can be achieved by Direct Numerical Simulation (DNS) with nowadays computers. 
One way around this difficulty is to resort to Large Eddy Simulations (or LES, 
see e.g. \cite{lesieur_rev,piomelli,meneveaukatz,sagaut} and references therein). Such techniques are widely 
used in engineering contexts, as well as in atmospheric sciences and, to a lesser extent, in geophysics 
(see \cite{vincent}) and astrophysics.
Another class of models is based on two-point closures, like the Eddy Damped Quasi Normal Markovian 
approximation, or EDQNM \cite{sao}.  These models, developed in the mid seventies, gave rise to successful 
LES when taking their eddy viscosity formulation \cite{chollet_lesieur,lesieur_book}. 
Such LES techniques have been used mostly in conjunction with pseudo-spectral methods, 
since being best expressed in Fourier space in terms of energy spectra. 

In this paper, we propose a new LES formulation that generalizes the usual EDQNM approach which is based on 
a Kolmogorov $k^{-5/3}$ spectrum (K41 hereafter), by allowing for a priori any kind of energy spectrum 
as may occur in the complex dynamical evolution of various turbulent flows, since our method (as will be shown later) 
is based on the evaluation of the transfer terms for energy and helicity. 
For example, there are small intermittency corrections to the K41 spectrum due to the 
presence of strong localized vortex filament structures in fluid turbulence; similarly, 
the presence of waves may alter the energy spectrum (see, e.g., \cite{galtier_waves}).
The method proposed here may also be particularly important 
when dealing with magnetohydrodynamics (MHD) flows, i.e. when coupling the velocity to the magnetic induction. 
In that case, the energy spectra can be either shallower \cite{grappin_05} or steeper than $k^{-5/3}$, 
because of anisotropy induced by a uniform magnetic field leading to Alfv\'en wave propagation and
to weak turbulence for strong magnetic background \cite{galtier_weak}. 
Similarly, in the case of strong correlations between velocity and magnetic fields 
\cite{grappin}, spectra that differ from the classical K41 phenomenology may emerge.
Note that, although we focus here on neutral flows, the extension of our model to
conductive MHD flows presents no particular difficulties \cite{MHDpap}.

In general, the traditional formulation of turbulent energy transfers only takes into account the energy of the flow 
but not its helicity. 
However, the closure transfer terms for helicity are well known in the helical case \cite{lesieur_book}, 
including in MHD \cite{PFL}. 
The kinetic helicity $H= 1/2<\textbf{v} \cdot \textbf{w}>$ (where $\textbf{w}= \nabla \times {\textbf{v}}$ is the vorticity) 
represents the lack of invariance of the flow by plane symmetry ($\textbf{w}$ is an axial vector). 
This global invariant of the Euler equation \cite{moffatt_tsinober} has been little studied until recently 
(see however \cite{lesieur_book,waleffe,holm_kerr,chen_eyink} and references therein).
Furthermore, the intermittent structures that populate a turbulent flow at small scales, 
namely the vortex filaments, are known to be helical (see \cite{moffatt_tsinober,holm_kerr,chen_eyink}); 
this implies that the nonlinear transfer terms involving small scales are weakened. 
This is consistent with several recent findings, namely:
(i) helical vortex tubes, in a wavelet decomposition of a turbulent flow into a Gaussian component and a structure component, 
represent close to 99\% of the energy and corresponds to the strong tails of the probability distribution 
function of the velocity gradients \cite{farge_01};
(ii) in a decomposition of the velocity field into large V and small v components, dropping (artificially) the nonlocal 
(in scale) nonlinear interactions (vV) leads to less intermittency \cite{LDN01}, indicating that intermittency 
involves interactions between structures (like vortex tubes) that incorporate small scales and large (integral) scales through a large aspect ratio;
and
(iii) the spectrum of helicity is close to $k^{-5/3}$ in the K41 range for energy, but not quite: the relative helicity
$ \tilde \rho = H(k)/{kE(k)} $
decreases more slowly than 1/k \cite{pablo_hel,alex_long}, indicating that the return to full isotropy is not 
as fast as one may have conjectured in the small scales.
Finally, helicity is also invoked as possibly responsible for the so-called bottleneck effect, i.e. 
the accumulation of energy at the onset of the dissipation range \cite{kurien}, although it 
is not clear whether this effect is or not an inertial range phenomenon \cite{alex_long}.

Our dynamical spectral LES model, based on the EDQNM closure, is described in Section II.
In Section III, numerical tests of the model are performed by comparisons with three-dimensional direct numerical 
simulations (DNS) for strong helical ABC flows \cite{dombre}, as well as a Chollet-Lesieur approach \cite{chollet_lesieur}.
Predictions for high Reynolds number flows are also given.
Section IV is the conclusion. 
Finally, details on closure expressions of the nonlinear transfers for energy and
helicity, and on the numerical implementation of the model are respectively given in Appendix A and B.
\vspace{10pt}
\section{Model description}
\subsection{Equations}
Let us consider the Fourier transform of the velocity $\textbf{v}(\textbf{x},t)$  and the
vorticity $\textbf{w}(\textbf{x},t)=\nabla \times \textbf{v}(\textbf{x},t)$ fields at wavevector $\textbf{k}$:
 \begin{equation}
   \textbf{v}(\textbf{k},t) = \int_{-\infty}^\infty \textbf{v}(\textbf{x},t) e^{-i\textbf{k}.\textbf{x}} \textbf{dx}
\end{equation}
 \begin{equation}
   \textbf{w}(\textbf{k},t) = \int_{-\infty}^\infty \textbf{w}(\textbf{x},t) e^{-i\textbf{k}.\textbf{x}} \textbf{dx}.
\end{equation}
In terms of the Fourier coefficients of the velocity components, the Navier-Stokes equation for an incompressible flow, 
with constant unit density, reads: 
\begin{equation}
(\partial_t + \nu k^2 )v_\alpha(\textbf{k},t) = 
 t_\alpha^v(\textbf{k},t) + F_\alpha^v(\textbf{k},t)\ ,\label{NSf}
\end{equation}
where $\bf{F}^v$ is the driving force, $\nu$ is the kinematic viscosity and $\textbf{t}^v(\textbf{k},t)$ 
is a bilinear operator written as:  
\begin{equation}
t_\alpha^v(\textbf{k},t)=-iP_{\alpha \beta}(\textbf{k})k_ \gamma\sum_{\textbf{p}+ \textbf{q} = 
\textbf{k}}v_{\beta}(\textbf{p},t)v_{\gamma}(\textbf{q},t) \ ;
\end{equation}
$P_{\alpha \beta}(\textbf{k})=\delta_{\alpha\beta}-k_\alpha k_\beta/k^2$ is the projector on solenoidal vectors.
In the absence of viscosity, both the total kinetic energy $E=1/2<{\textbf{v}}^2>$ and the total helicity 
$H= 1/2<\textbf{v} \cdot \textbf{w}>$ are conserved and they are thus thought to play an important dynamical role 
in the temporal evolution of the fluid. The direct cascade of energy to the small scales and the related 
cascade of helicity \cite{brissaud} stem from these conservation laws. 
Furthermore, because helicity is thought to play an important role in the small scales (see e.g.  \cite{moffatt_tsinober} 
and references therein and more recently \cite{farge_01,holm_kerr,kurien}), we are taking in this paper the approach 
of following the time evolution of both the energy and helicity spectra (see below).
Taking the rotational of Eq. (\ref{NSf}) in Fourier space leads to:
\begin{equation}
(\partial_t + \nu k^2 )w_\alpha(\textbf{k},t) = 
t_\alpha^w(\textbf{k},t)+ F_\alpha^w(\textbf{k},t)\ . \label{NSfw}
\end{equation}
with 
\begin{eqnarray}
t_\alpha^w(\textbf{k},t) & = & \varepsilon_{\alpha \delta \beta} \ k_\delta k_\gamma \!\! \sum_{\textbf{p}+ \textbf{q}=
\textbf{k}}v_{\beta}(\textbf{p},t)v_{\gamma}(\textbf{q,t}) \ , \\
F_\alpha^w(\textbf{k},t) & = & i \ \varepsilon_{\alpha \delta \beta} \ k_\delta \ F_\beta^v(\textbf{k},t) \ .
\end{eqnarray}

We respectively define the modal spectra of energy  $\mathcal{E}(\textbf{k},t)$ 
and helicity  $\mathcal{H}(\textbf{k},t)$  in the usual way as:
\begin{equation}
   \mathcal{E}(\textbf{k},t) = \frac{1}{2}\textbf{v}(\textbf{k},t) \cdot \textbf{v}^*(\textbf{k},t) \ ,
\end{equation}
\begin{equation}
   \mathcal{H}(\textbf{k},t) = \frac{1}{2}\textbf{v}(\textbf{k},t) \cdot \textbf{w}^*(\textbf{k},t) \ ,
\end{equation}
where stars stand for complex conjugates. Note that ${\bf w}$ is a pseudo (axial) vector, and 
correspondingly the helicity is a pseudo scalar.
The integration of $\mathcal{E}(\textbf{k},t)$ and $\mathcal{H}(\textbf{k},t)$ over shells of 
radius $k=|\textbf{k}|$ respectively gives the isotropic energy  
$E(k,t)$ and helicity $H(k,t)$ spectra. Their spatio-temporal evolutions obey the following equations:
\begin{eqnarray}
(\partial_t + 2\nu k^2 )E(k,t)& = &
T_{E}(k,t) +  F_{E}(k,t) \label{energy}\\
(\partial_t + 2\nu k^2 )H(k,t)& = &
T_{H}(k,t) +  F_{H}(k,t)  
\label{helicity}
\end{eqnarray}
where $T_{E}(k,t)$ and $T_{H}(k,t)$ denote energy and helicity nonlinear transfers at wavenumber $k$. 
They are functionals of the tensors involving triple correlations between 
$\textbf{v}(\textbf{k},t)$, $\textbf{v}(\textbf{p},t)$ and $\textbf{v}(\textbf{q},t)$ with the constraint that 
$\textbf{p} + \textbf{q} = \textbf{k}$ due to the convolution term in Fourier space 
emanating from the nonlinearities of the primitive Navier-Stokes equation. 

Finally, under the closure hypothesis customary to the EDQNM approach (see \cite{lesieur_book} and references therein), 
the time evolution of $E(k,t)$ and $H(k,t)$ can be described by the EDQNM equations where the 
exact transfer terms $T_{E}(k,t)$ and $T_{H}(k,t)$ in the equations above are replaced by the closure 
evaluations denoted as $\widehat T_{E}(k,t)$ and $\widehat T_{H}(k,t)$. 
The closure is done at the level of fourth-order correlators which are expressed in terms of third-order ones, 
with a proportionality coefficient -- dimensionally, the inverse of a time -- taken as the sum of all 
characteristic rates appearing in a given problem, namely the linear (dispersive), nonlinear and dissipative rates. 
Tested against DNS \cite{lesieur_book}, these closures allow for exponential discretization in Fourier space and 
hence for exploration of high Reynolds number regimes. Their drawback is that all information above 
second-order moments is lost and phase information among Fourier modes is lost as well, so that, 
for example, intermittency is not present in this approach, nor are spatial structures. 

The full formulation of the EDQNM closure leads to a set of coupled integro-differential equations 
for the energy and helicity spectra $E(k,t)$ and $H(k,t)$, with the nonlinear transfers decomposed 
into emission terms ($S_{E_{1}}$, $S_{E_{3}}$ and $S_{H_{1}}$, $S_{H_{3}}$), 
and absorption terms ($S_{E_{2}}$, $S_{E_{4}}$ and $S_{H_{2}}$, $S_{H_{4}}$); note that we use
$S_{E_i}$ and $S_{H_i}$, with $i\in [1,4],$ as short-hand notations for the full spectral functions
$S_{E_i}(k,p,q,t)$ and $S_{H_i}(k,p,q,t)$. 
The expressions of these closure transfer terms are given in Appendix A.
Note that absorption terms are linear in the spectra, the dynamical evolution of which we are seeking, 
whereas emission terms are inhomogeneous terms involving the $p,q$ wavenumbers on which the double 
sum is taken (with $\textbf{p} + \textbf{q} = \textbf{k}$). 
The absorption term,  $S_{E_{2}}$, leads to the classical concept of eddy viscosity, 
whereas the emission term, $S_{E_{1}}$, is in general modeled as an eddy noise, although it is known 
through both experiments and DNS that the small scales are far from following a Gaussian distribution, 
with substantial wings corresponding to strong localized structures. Here, we
present a different and novel method to treat the emission term.

\vspace{10pt}
\subsection{Spectral filtering}
When dealing with an LES method, as a complement in the unresolved small scales to the dynamical evolution of 
the large scales following the Navier-Stokes equation, we need to partition Fourier space into three regions. 
This means that we need to introduce a buffer region between the scales that are completely resolved (above $k_c^{-1}$, 
where $k_c$ is a cut-off wavenumber depending on the resolution of the LES run), and the scales that are completely 
unresolved, say beyond $ak_c$ with $a$ of $\mathcal O (1)$. 
Following \cite{kraichnan} and according to the Test Field Model closure, 
the contribution of subgrid scales, to the explicitely resolved inertial 
scales, leads to an eddy viscosity depending both on the wavenumber and the energy spectrum at 
that wavenumber. 
It is also shown that beyond $2k_c$, 85\% of the transfer is covered by the eddy viscosity, 
while beyond $3k_c$, about 100\% is covered; we thus choose to take $a=3$.

More specifically, the truncation of equations (\ref{NSf}), (\ref{NSfw}), (\ref{energy}) and 
(\ref{helicity}) at two different wavenumbers $k=k_c$ and $k=3k_c$ gives
rise to three types of transfer terms, corresponding respectively to local, nonlocal and highly nonlocal interactions 
(where locality refers to Fourier space, i.e. interactions between modes of comparable wavenumber):

\noindent
(i) the fully resolved transfer terms $T^<_{E}$ and $T^<_{H}$ involve triadic interactions 
such that $k$, $p$, and $q$ are all three smaller than $k_c$; this interval is denoted $\Delta^<$;

\noindent
(ii) the intermediate nonlocal tranfer terms $T^>_{E,H}$, in which $p$ and/or $q$ are 
contained in the buffer zone between $k_c$ and $3k_c$ (hereafter denoted $\Delta^>$); and 

\noindent
(iii) the highly nonlocal tranfer terms $T^{>>}_{E,H}$,  in which $p$ and/or $q$ are larger than $3k_c$
(hereafter denoted $\Delta^{>>}$).

We choose  to model  $T^>_{E,H}$ and $T^{>>}_{E,H}$ in Eqs. (\ref{energy}) and (\ref{helicity}) by 
appropriately modified EDQNM transfer terms. 
We therefore need to know the behavior of both energy and helicity spectra after the cut-off wavenumber 
$k_c=N/2-1$, where $N$ is the linear grid resolution of the numerical simulation. 
Whereas it is customary to assume a $k^{-5/3}$ Kolmogorov spectrum in this intermediate range, 
here we choose a different approach, 
namely, between $k=k_c$ and $k=3k_c$, both spectra are assumed to behave as power-laws 
(with unspecified spectral indices) followed by an exponential decrease, viz.:
\begin{eqnarray}
E(k,t) & = & E_0 k^{-\alpha_E} e^{-\delta_E k}, \quad  k_c\le k<3k_c \label{fit_E}\\
H(k,t) & = & H_0 k^{-\alpha_H} e^{-\delta_H k}, \quad  k_c\le k<3k_c\label{fit_H} 
\end{eqnarray}
where $\alpha_E$, $\delta_E$, $E_0$, and $\alpha_H$, $\delta_H$, $H_0$ are evaluated, at each
time step, by a mean square fit of the energy and helicity spectra, respectively. 
Note that it is understood that the Schwarz inequality $|H(k)|\le kE(k)$ is fulfilled at all times.
When either $\delta_E$ or $\delta_H$ is close to zero, we consider that the energy (or helicity) spectrum 
has an infinite inertial range with a $k^{-\alpha_{E,H}}$ power law (see eq. (\ref{annick1})), so we can write: 
\begin{eqnarray}
E(k,t) & = & E_0 k^{-\alpha_E}, \quad  3k_c\le k<\infty\label{fit_E_sup} \ , \\
H(k,t) & = & H_0 k^{-\alpha_H}, \quad  3k_c\le k<\infty\label{fit_H_sup} \ .
\end{eqnarray}

\vspace{10pt}
\subsection{Eddy viscosity}
In the context of spectral models for the Navier-Stokes equation,
the concept of eddy viscosity was introduced by Kraichnan \cite{kraichnan}.
This transport coefficient, denoted $\nu(k|k_c,t)$, allows to model the nonlinear transfer through a dissipative 
mechanism, as first hypothesized by Heisenberg. 
With $T^>_{E}$ and $T^{>>}_{E}$ terms defined above, and where the hat denotes the fact that 
the EDQNM formulation of these partial transfers is taken, the eddy viscosity reads:
\begin{eqnarray}
\nu(k|k_c,t) &=&-\frac{\widehat T_E^>(k,t)+\widehat T_E^{>>}(k,t)}{2k^2E(k,t)} \nonumber  \\ 
&=&  \nu^> (k|k_c,t) + \nu^{>>}(k|k_c,t) \ ,
\end{eqnarray}
thus separating the contribution stemming from the buffer zone ($\Delta^>$) and from the outer zone ($\Delta^{>>}$). 
Note that only the part of the transfer proportional to the energy spectrum at 
wavenumber $k$ (i.e. the $S_{E_2}(k,p,q,t)$ part defined in Appendix A) is taken 
into account in the derivation of $\nu^> (k|k_c,t)$.
Indeed, in our model, the closure transfer term $\widehat{T}_{E}^{>}(k,t)$ is integrated at each time step, 
but an eddy viscosity from this whole transfer term cannot be extracted; only the part that explicitely 
contains $E(k,t)$ (the linear part of the transfer in E(k,t)) enables the derivation 
of an eddy viscosity, namely:
\begin{eqnarray}
\nu^{>}(k|k_c,t)& = & - \iint_{\Delta^>}\frac{\theta_{_{kpq}}S_{E_2}(k,p,q,t)}{2 k^2E(k,t)}dpdq
\nonumber \\
& = &\iint_{\Delta^>}\theta_{_{kpq}}\frac{p^2}{2k^2q}(xy+z^3)E(q,t)dpdq
\nonumber
\end{eqnarray}
Let us now evaluate the eddy viscosity in the outer region, $\nu^{>>}(k|k_c,t)$, coming from
the highly nonlocal EDQNM transfer terms. 
Since the ($k$,$p$,$q$) triangles are very elongated in the $\Delta^{>>}$ zone, with $k\ll  p,q$, 
an algebraical simplification occurs leading to an explicit expression for $\nu^{>>}(k|k_c,t)$. 
Indeed, it has been shown \cite{metais_lesieur92} that a Taylor expansion of $\widehat {T}_E^{>>}(k,t)$ with respect to $k/q$
leads, at first order, to the following asymptotic transfer term:
\begin{equation}
\widehat {T}_{E}^{>>}(k)=-\frac{2}{15}k^2E(k)\int_{3k_c}^{\infty}\theta_{_{kpp}}[5E(p)+p\frac{\partial E(p)}{\partial p}]dp \ ,
\label{ener} 
\end{equation}
where time dependency is omitted for simplicity.
Since now $\widehat {T}_{E}^{>>}(k)$ explicitely depends on $E(k)$, it is straightforward to formulate the
corresponding eddy viscosity $\nu^{>>}(k|k_c,t)$ thus defined as:
\begin{equation}
\nu^{>>}(k|k_c,t)=\frac{1}{15}\int_{3k_c}^{\infty}\theta_{_{kpp}}[5E(p)+p\frac{\partial E(p)}{\partial p}]dp \ .
\end{equation}
When $E(p)$ is replaced by its power law-exponential decay approximation (see Eq. (\ref{fit_E_sup})),
we recover the so-called ``plateau-peak'' model
\cite{lesieur_book}:
\begin{equation}
\nu^{>>}(k|k_c,t)\simeq 0.31\frac{5-\alpha_E}{1-\alpha_E}\sqrt{3-\alpha_E}C_K^{-\frac{3}{2}}
\Big[\frac{E(3k_c,t)}{3k_c}\Big]^\frac{1}{2}.
\label{annick1}
\end{equation}
Finally, in the energy equation Eq. (\ref{energy}), the total eddy viscosity derived from the nonlocal and highly 
nonlocal transfer terms is simply obtained by adding the two contributions, as stated before:
$\nu(k|k_c,t)=\nu^{>}(k|k_c,t) + \nu^{>>}(k|k_c,t)$.

Note that, in the helicity equation Eq. (\ref{helicity}), the transport coefficient stemming from the helicity 
transfer term $T_H(k,t)$ can be similarly evaluated in the buffer zone and the outer zone, and written as:
\begin{eqnarray}
\nu_H(k|k_c,t) &=&-\frac{\widehat T_H^>(k,t)+\widehat T_H^{>>}(k,t)}{2k^2H(k,t)} \nonumber  \\ 
&=&  \nu_H^> (k|k_c,t) + \nu_H^{>>}(k|k_c,t) ,
\end{eqnarray}
with 
\begin{equation}
\nu_H^{>}(k|k_c,t) =  - \iint_{\Delta^>}\frac{\theta_{_{kpq}}S_{H_2}(k,p,q,t)}{2 k^2H(k,t)}dpdq. 
\end{equation}
It is straighfoward to show that this eddy viscosity part has the same formulation than $\nu^{>}(k|k_c,t)$
(see Appendix A for $S_{H_2}(k,p,q,t)$ definition). 
For the helicity transfer term $\widehat {T}_H^{>>}(k,t)$, simple algebraic calculations lead to:
\begin{equation}
\widehat {T}_H^{>>}(k)=-\frac{2}{15}k^2H(k)\int_{3k_c}^{\infty}\theta_{_{kpp}}[5E(p)+p\frac{\partial E(p)}{\partial p}]dp. 
\label{hely}
\end{equation}
The integrands in Eqs. (\ref{ener}) and  (\ref{hely}) are thus identical; this in turn provides the same eddy viscosity
in the outer domain than for the energy, namely $\nu_H^{>>}(k|k_c,t)=\nu^{>>}(k|k_c,t)$. 
Altogether, the same total eddy viscosity appears in both the energy and helicity equations.
This is expected from the formulation of the spectral closure, in which the temporal dynamics of the 
second-order velocity correlation 
function is separated into its symmetric (energetic) and anti-symmetric (helical) parts.

\vspace{10pt}
\subsection{Helical eddy diffusivity}
At wavenumber $k$, the energy transfer obtained from the use of the EDQNM closure
involves a linear term in the helicity spectrum $H(k,t)$  (specifically, $S_{E_4}(k,p,q,t)$ defined in Appendix A, Eq. \ref{S_E}); 
from this term, a new transport coefficient, similar to the $\nu^{>}(k|k_c,t)$ eddy viscosity, can be built. 
In the buffer zone, this new coefficient, hereafter named ``helical eddy diffusivity'', reads:
\begin{eqnarray}
\widetilde \nu^>(k|k_c,t)& = &\iint_{\Delta^>}\frac{\theta_{_{kpq}}S_{E_4}(k,p,q,t)}{2 k^2H(k,t)}dpdq
\nonumber \\
& = &\iint_{\Delta^>}\theta_{kpq}\frac{1}{2k^2q}z(1-y^2)H(q,t)dpdq
\nonumber \\
\end{eqnarray}
Note that, dimensionally, this helical diffusivity $\widetilde \nu$ scales as $\nu / k$.
As before a total helical eddy diffusivity can be defined as 
$\widetilde \nu(k|k_c,t)=\widetilde \nu^>(k|k_c,t) + \widetilde \nu^{>>}(k|k_c,t)$,
where $\tilde \nu(k|k_c,t)$ represents the contribution of the 
small-scale helicity spectrum to the kinetic energy dissipation.
Recall that, in the outer zone, the Taylor expansion of the highly nonlocal transfer, $\widehat T_E^{>>}(k,t)$, 
with respect to $k/q \ll 1$, leads at first order to Eq. (\ref{ener}), with no linear contribution
from $H(k,t)$. We therefore assume that the transfer part associated with helical motions in the outer zone 
is negligible, such as $\widetilde \nu^{>>}(k|k_c,t)=0$.
The total helical eddy diffusivity thus reduces to $\widetilde \nu(k|k_c,t) = \widetilde \nu^>(k|k_c,t)$.
\vspace{10pt}
\subsection{Emission transfer terms}
The parts of the EDQNM transfer terms which are not included either in the eddy viscosity or 
in the helical eddy diffusivity, involve energy and helicity interactions at wavenumbers 
$p$ and $q$ both larger than $k_c$. Respectively denoted $\widehat {T}_E^ {\ pq}(k,t)$ 
and $\widehat {T}_H^{\ pq}(k,t)$, they read: 
\begin{eqnarray}
\widehat{T}_{E}^{\ pq}(k,t) & = &
\int_{k_c}^{3k_c}\!\!\!\int_{k-p}^{k+p}\!\!\! \theta_{_{kpq}}(t)\big(S_{E_1}+S_{E_3}\big)dpdq \nonumber \\
\widehat{T}_{H}^{\ pq}(k,t) & = &
\int_{k_c}^{3k_c}\!\!\!\int_{k-p}^{k+p}\!\!\!\theta_{_{kpq}}(t)\big(S_{H_1}+S_{H_3}\big)dpdq \nonumber \\
\end{eqnarray} 
where $S_{E_i,H_i}$ stands for $S_{E_i,H_i}(k,p,q,t)$.

On the one hand, the established eddy viscosity and helical eddy diffusivity can be directly used
in the Navier-Stokes equation for the modal energy and helicity spectra, 
$\mathcal{E}(\textbf{k},t)$ and $\mathcal{H}(\textbf{k},t)$ respectively.
On the other hand, in order to implement in these modal equations, the isotropic transfers 
$\widehat{T}_{E}^{\ pq}(k,t)$ and $\widehat{T}_{H}^{\ pq}(k,t)$, we assume that they are uniformly distributed
among all ${\bf k}$ wavevectors belonging to the same k-shell. This means that the nonlocal modal energy and helicity
transfers, respectively $\widehat{\mathcal{T}}_E^{pq}(\textbf{k},t)$
and $\widehat{\mathcal{T}}_H^{pq}(\textbf{k},t)$, can be expressed as
$\mathcal{\widehat{T}}_E^{pq}(\textbf{k},t)=\widehat{T}_E^{pq}(k,t)/4\pi k^2$ 
and $\mathcal{\widehat{T}}_H^{pq}(\textbf{k},t)=\widehat{T}_H^{pq}(k,t)/4\pi k^2$. 

\vspace{10pt}
\subsection{Numerical field reconstruction}
To compute our LES model for all $k<k_c$, we proceed in two steps.
At a given time, the Navier-Stokes equation is first solved using the eddy viscosity 
and the helical eddy diffusivity, namely:
\begin{eqnarray}
(\partial_t + \nu k^2 )v_\alpha(\textbf{k},t) 
&=&-iP_{\alpha \beta}(\textbf{k})k_ \gamma\!\!\!\sum_{\substack{\textbf{p}+ \textbf{q} = 
\textbf{k} \\ k,p,q<k_c}}\!\!\!v_{\beta}(\textbf{p},t)v_{\gamma}(\textbf{q}) \nonumber\\
& & - \nu(k|k_c,t)k^2 v_\alpha(\textbf{k},t) \nonumber \\
& & - \widetilde \nu(k|k_c,t)k^2 w_\alpha(\textbf{k},t) \nonumber \\
& & + F_\alpha^v(\textbf{k},t)\ .
\label{NSfmodel}
\end{eqnarray}

Then, the effects of the emission terms, $\widehat{T}_E^{\ pq}(k,t)$ and $\widehat{T}_H^{\ pq}(k,t)$, 
are introduced in the numerical scheme. 
In most previous studies, these terms are taken into account through a random force, 
uncorrelated in time (see e.g. \cite{Chasnov}); this corresponds to the vision that they represent 
an eddy noise originating from the small scales. However, the small scales are all but uncorrelated noise; 
the phase relationships within the small-scale structures play an important role, albeit not fully understood,
in the flow dynamics. It is well-known that a random field with a  $k^{-5/3}$ Kolomogorov energy spectrum, 
but otherwise random phases of the Fourier coefficients, is very different from an actual turbulent flow, 
lacking, in particular, the strong vortex tubes so prevalent in highly turbulent flows. 
Similarly, it has been recently shown \cite{nagoya_alex} that, upon phase randomization, 
the ratio of nonlocal energy transfer (i.e. the transfer involving widely separated scales) 
to total energy transfer reduces to a negligible amount, whereas this ratio is close to 20 \%
at the resolutions of the performed numerical experiments, corresponding to a Taylor Reynolds number of about $10^3$.
These considerations lead us to directly incorporate the emission terms 
in the second step of the our numerical procedure. The modal spectra of the energy and the helicity,
associated to the ${\bf v}(\textbf{k},t)$ field computed from Eq. (\ref{NSfmodel}),
now has to verify the following equations where the emission transfer terms are taken into account; 
\begin{eqnarray}
\big
(\partial_t +  2\nu k^2 \big)\mathcal{E}(\textbf{k},t) & = &-2\nu(k|k_c,t) k^2\mathcal{E}(\textbf{k},t) \nonumber \\ 
& & -2\widetilde \nu(k|k_c,t) k^2\mathcal{H}(\textbf{k},t) \nonumber \\
& & + \mathcal{T}_E^<(\textbf{k},t) + \frac{\widehat{T}_E^{pq}(k,t)}{4\pi k^2} \nonumber \\
& & + \mathcal{F}_{E}(\textbf{k},t), \label{be-system} \\     
{\big
(\partial_t +  2\nu k^2 \big)\mathcal{H}(\textbf{k},t)} & = &
-2 \nu(k|k_c,t) k^2\mathcal{H}(\textbf{k},t) \nonumber \\ 
& & -2 \widetilde \nu(k|k_c,t)k^4\mathcal{E}(\textbf{k},t) \nonumber \\
& & + \mathcal{T}_H^<(\textbf{k},t) + \frac{\widehat{T}_H^{pq}(k,t)}{4\pi k^2} \nonumber \\
& & + \mathcal{F}_{H}(\textbf{k},t). 
\label{bh-system}
\end{eqnarray}
where $\mathcal{F}_{E}(\textbf{k},t)$ and $\mathcal{F}_{H}(\textbf{k},t)$
denote the spectral terms stemming from the driving force. 
Recall that $\mathcal{T}_E^<(\textbf{k},t)$ and $\mathcal{T}_H^<(\textbf{k},t)$ are the resolved transfer
terms based on triadic velocity interactions with $k$, $p$, and $q$ all smaller than $k_c$.
Once the uptaded $\mathcal{E}(\textbf{k},t)$ and $\mathcal{H}(\textbf{k},t)$ modal spectra are obtained, 
the velocity field is updated.
However, a difficulty immediately arises: the phase relationships between the three components 
of the velocity field in the EDQNM (and other) closures is of course a priori lost. 
We thus proceed to the reconstruction of the three spectral velocity components,
written as $v_\alpha(\textbf{k},t)=\rho_\alpha(\textbf{k},t) e^{i\phi_\alpha(\textbf{k},t)}$, and  
rebuild the different velocity phases by using the incompressibility and
realisability ($|\mathcal H({\bf k},t)|\le k \mathcal E({\bf k},t)$) conditions, as explained in Appendix B. 
\section{Numerical tests of the model}
\vspace{10pt}
\subsection{Numerical setup}
In order to assess the model accuracy to reproduce the physics involved in fluid flows, we performed 
Direct Numerical Simulations (DNS) of the Navier-Stokes equation and computations using our 
LES formulation. We denote LES P the code with partial recovery of phases and without helical effects
(i.e., with $\tilde \nu \equiv 0$ and $\widehat T_H^{pq}\equiv 0$), 
and LES PH the code with helical effects incorporated. 
In our LES description, the energy spectra - and helicity spectra when considered - of the 
subgrid scales self adapt to the large scale resolved spectra ({\it i.e.} no spectral scaling laws are prescribed). 
We can therefore study a variety of flows, such as either low or high Reynolds number flows, 
or the early phases of the temporal development of flows when Kolmogorov spectra are not yet established.
Indeed, the cut-off wavenumber, $k_c$, can as well lie in the dissipation range instead of the inertial
range which is the case of standard LES approaches based on closures together with a $k^{-5/3}$ Kolmogorov spectrum. 
This enables accurate comparisons with DNS at a given viscosity.
We also compared our numerical approach to a Chollet-Lesieur LES model
(CL) \cite{chollet_lesieur}, where a $\nu=2.e^{-3}$ kinematic viscosity is added to 
the turbulent viscosity for comparison purpose (see Table~\ref{table1}).
The codes use a pseudo-spectral Fourier method in a $[0-2\pi]^3$-periodic box and 
an Adams-Bashforth second-order scheme in time.

To test the ability of our LES models to reproduce helical flow features, 
we focus on flows driven by a prototype Beltrami flow (${\bf v}=\pm {\bf w}$), namely the ABC flow (see e.g. \cite{dombre}):
\begin{equation} 
{{\bf  F}_{\rm ABC}(k_0)}=   \left[ 
\begin{array}{c} 
B\cos(k_0 y)+C\sin(k_0 z) \\ 
C\cos(k_0 z)+A\sin(k_0 x)\\ 
 A\cos(k_0 x)+B\sin(k_0 y)  
\end{array} \right] , \label{eq:Fabc} \end{equation} 
with $k_0=2$, and $A=B=C=1$.

The run parameters are summarized in Table~\ref{table1}. 
The definitions used for the integral scale, $L$, and the Taylor microscale, $\lambda$,
are based on the kinetic energy spectrum  $E(k)$; 
respectively $L = 2\pi \int{k^{-1} E(k) dk}/ \int{E(k) dk}$ and
$\lambda = 2\pi \left[\int{E(k) dk}/ \int{k^2 E(k) dk}\right]^{1/2}$.
Note that the characteristic flow quantities - $L$ and $\lambda$ scales, \rms\  velocity, $U_{rms}$,
nonlinear time scale, ${\tau_{_{NL}}}$, and Reynolds number, ${\Rv}$ - are time averaged quantities
once the steady state is achieved for the computed flow.
\begin{table*}
\caption{\label{table1}Parameters of the simulations {\bf I} to {\bf IX}. 
Linear grid resolution $N$, kinematic viscosity $\nu$
and time averaged quantities:
Taylor microscale $\lambda$ and integral scale $L$;
\rms velocity $U_{rms}=<{\bf v}^2>^{1/2}$;
integral Reynolds number $\Rv=$$\ U_{rms}L/\nu$; eddy turnover time $\tau_{NL}=L/U_{rms}$;
$t_{M}$ is the final time of integration.
Note that the {\bf Ir} label stands for $64^3$ reduced data obtained from the $256^3$
DNS computation. The LES P (vs. PH) label stands for our model computations
without (resp. with) incorporating
the helicity transport coefficients $\tilde \nu$ and emission transfer terms $\widehat T_H^{pq}$.
The LES CL label stands for a Chollet-Lesieur scheme where the kinematic viscosity is added to
the scheme eddy viscosity
(see text).}
\begin{ruledtabular}
\begin{tabular}{cccccccccc}
& & $N$ & $\nu$ & ${\lambda}$ & $L$ & $U_{rms}$& ${\Rv}$ & ${\tau_{_{NL}}}$ &  $t_{M}$\\
{\bf I}   & DNS & $256$ & $5.e^{-3}$ & $0.81$ & $2.38$ & $3.19$  &  $1525$ & $0.75$ & $60$ \\
{\bf Ir}   & Reduced DNS  & $64$ & $5.e^{-3}$ & $0.92$ & $2.38$  & $3.19$  & $1530$  & $0.75$ & $60$ \\
{\bf II} & LES PH & $64$& $5.e^{-3}$ & $0.93$ & $2.37$ & $3.20$  &  $1519$ & $0.74$ & $60$ \\
{\bf III} & LES P& $64$ & $5.e^{-3}$ & $0.93$ & $2.40$ & $3.22$  &  $1544$ & $0.74$ & $60$ \\
{\bf IV}   & DNS & $512$ & $2.e^{-3}$ & $0.49$ & $2.32$ &  $3.34$ &  $3881$ & $0.70$ & $7.0$ \\
{\bf V}   & LES PH & $128$ & $2.e^{-3}$ & $0.59$ & $2.30$ & $3.36$  &  $3877$ & $0.69$ & $7.0$ \\
{\bf VI}   & LES P & $128$ & $2.e^{-3}$ & $0.59$ & $2.33$ & $3.37$  &  $3925$ & $0.69$ & $7.0$ \\
{\bf VII}   & LES CL & $128$ & $2.e^{-3}$ & $0.66$ & $2.38$ & $3.29$  &  $3928$ & $0.72$ & $7.0$ \\
{\bf VIII}  & LES PH & $256$ & $5.e^{-4}$ & $0.36$ & $2.47$ & $3.38$  &  $16693$ & $0.73$ & $10$ \\
{\bf IX}  & LES P & $256$ & $5.e^{-4}$ & $0.36$ & $2.35$ & $3.31$  &  $15565$ & $0.71$ & $10$ \\
\end{tabular}
\end{ruledtabular}
\end{table*}
\vspace{10pt}
\subsection{Spectral features}
\begin{figure}[h!]
   \centerline{\epsfig{file=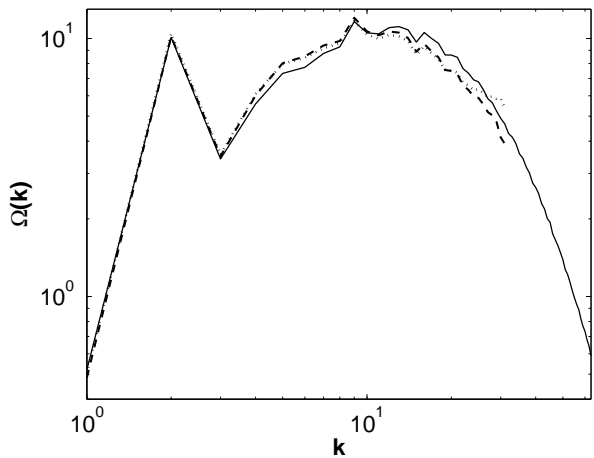, width=\linewidth}}
    \caption{Time averaged enstrophy spectra for runs {\bf I} (solid line), 
{\bf II} (dashed line) and {\bf III} (dotted line). }  
\label{Enstrophie_spectrum}  
\end{figure}

We first investigate the flow spectral behavior on one-dimensional energy, enstrophy and helicity spectra 
obtained from the different models. These spectra are averaged over $67$ nonlinear turnover times 
spanning the flow steady phase from $t=10.0$ up to $t=60.0$, the final time reached in the simulation.
\begin{figure}[h!]
   \centerline{\epsfig{file=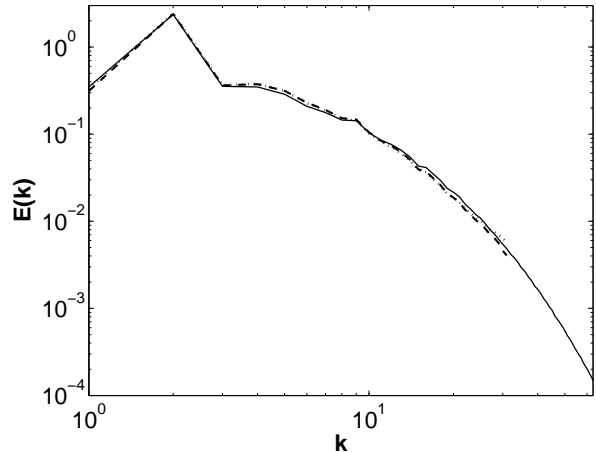,
   width=\linewidth}}
     \caption{Time averaged energy spectra $<E(k,t)>$ for data {\bf I} ($256^3$ DNS, solid line), 
{\bf II} ($64^3$ LES PH, dashed line) and {\bf III} ($64^3$ LES P, dotted line). See Table~\ref{table1}.}
   \label{Energy_spectrum}
\end{figure}
\noindent
Fig. \ref{Energy_spectrum} shows the time averaged energy spectra $<E(k,t)>$ for LES and DNS 
computations (runs {\bf I}, {\bf II} and {\bf III}).  
Both LES results show good agreement with the corresponding DNS ones, up to $k_c=31$, 
the maximum wavenumber of the LES calculations. Small differences are observed in both LES P and 
LES PH models at the largest wavenumbers.
When looking at the vorticity density spectra, these differences are amplified as
small scales are emphasized (see Fig.~\ref{Enstrophie_spectrum}). However, the mean characteristic 
wavenumber of the velocity gradients, defined as the maximum of the time averaged vorticity density spectrum, 
corresponds to $k=9$ in all runs.

Note that, for the different flows, the time averaged Reynolds number and characteristics integral scale 
are almost the same (see Table~\ref{table1}), while the Taylor microscale, ${\lambda}$,
and its associated Reynolds number, ${\Rla}$, differ due to non negligible intensities of the
enstrophy spectrum 
(namely, $k^2 E(k)$ used to compute $\lambda$, with isotropy assumed) 
after the cut-off wavenumber $k_{c}$.  
When down-sizing the $256^3$ DNS data to $64^3$ grid points by
filtering all wave vectors ${\bf k}$ such as $|{\bf k}| >  k_c$ (case {\bf Ir} in Table~\ref{table1}), 
the Taylor small scale quantities obtained from {\bf Ir}, LES PH {\bf II} and LES P {\bf III} runs get closer.
Thus, our LES models can estimate the mean correlation length scale of the vorticity when 
the smallest scales are properly filtered out. 

\begin{figure}[h!]
   \centerline{\epsfig{file=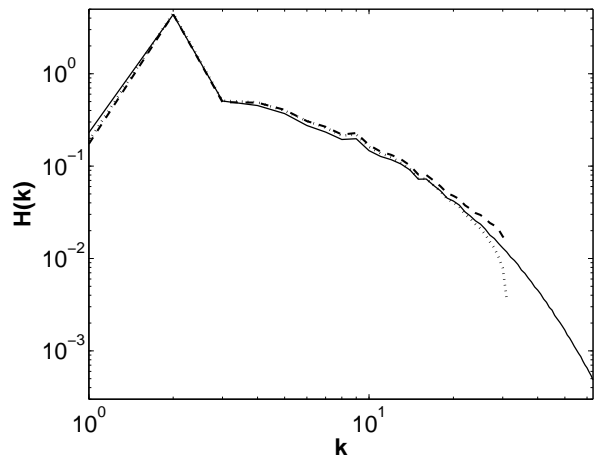,
   width=\linewidth}}
     \caption{Time averaged helicity spectra $<H(k,t)>$ for runs {\bf I} (solid line), 
{\bf II} (dashed line) and {\bf III} (dotted line). }
   \label{Helicity_spectrum}
\end{figure}
The time averaged helicity spectra are plotted in Fig.~\ref{Helicity_spectrum}.
At large scale, the LES P and LES PH models provide a close approximation of the DNS 
helicity spectrum up to $ k \sim 20$. 
For $k \geq 20$, the LES PH model, designed to take into account helical effects, slightly overestimates
the $<H(k,t)>$ magnitudes of DNS data, while the LES P model dissipates too much helicity.

\vspace{10pt}
\subsection{Temporal evolution}
In this section, to study the temporal behavior of the flows, in the spirit of the analysis performed for  
freely evolving fluids, we focus on the temporal phase 
before the steady state regime.
\begin{figure}[ht]
   \centerline{\epsfig{file=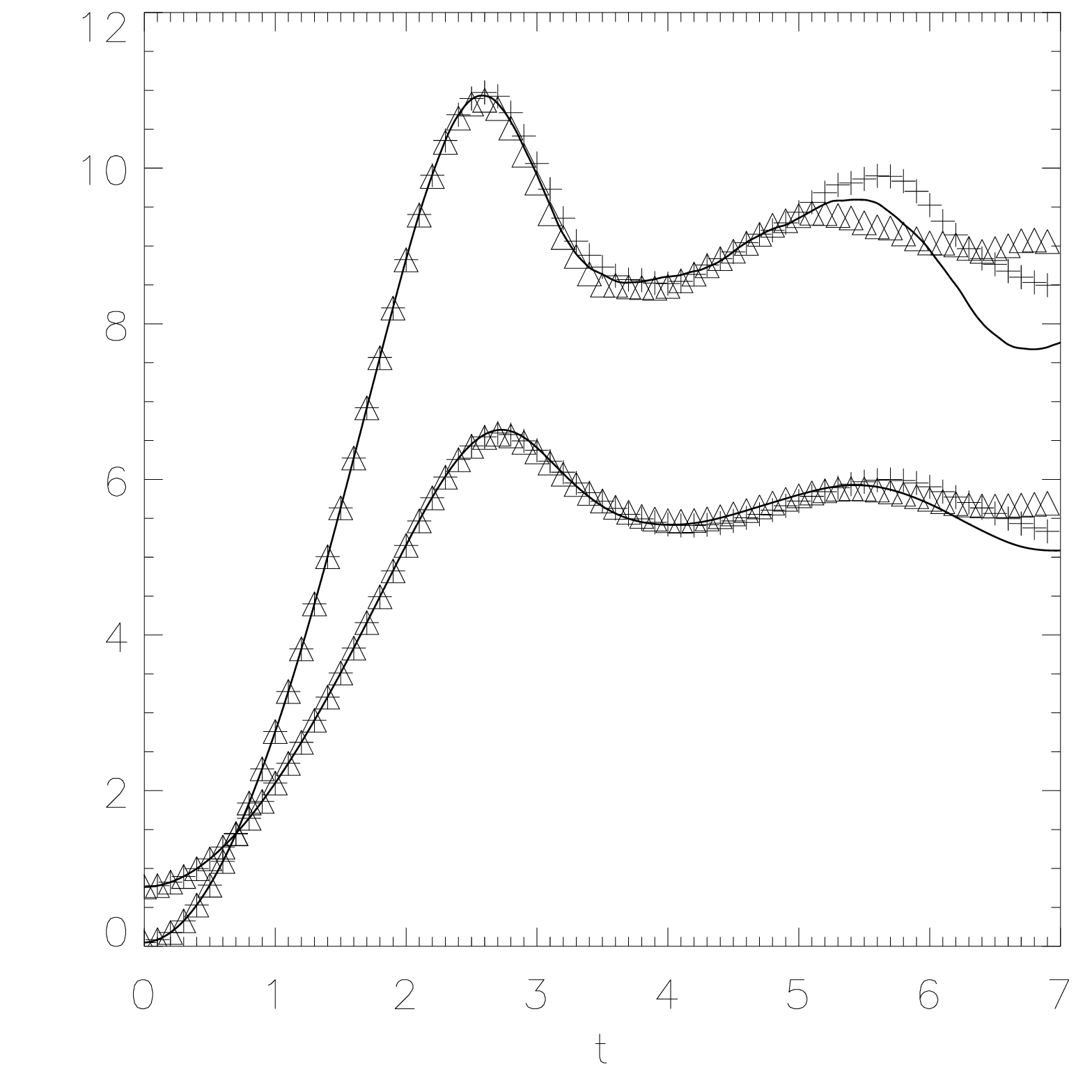,
   width=\linewidth,height=50mm}}
     \caption{Evolution of kinetic energy $E(t)$, lower curves, and helicity $H(t)$, upper curves,
for runs {\bf IV} ($512^3$ DNS, solid line), 
     {\bf V} ($128^3$ LES PH, plusses) and {\bf VI} ($128^3$ LES P, triangles).}
   \label{Energy_evolution}
\end{figure}
\noindent
Fig.~\ref{Energy_evolution} shows the evolution of the global kinetic energy  $E(t)$ 
and helicity $H(t)$ for DNS together  with LES computations, namely runs  
{\bf IV} ($512^3$ DNS), {\bf V} ($128^3$ LES PH) and {\bf VI} ($128^3$ LES P).
We first observe that for both LES models, with and without helical effects, energies 
closely follow the growth phase of the DNS energy. 
Indeed, during the inviscid phase, $t \leq 1.0$, the small scales are generated with negligeble effects 
on large scales, since their intensities are very weak. Thus, at that times, the LES modelling 
has only a reduced action.  Later on, during the following growth phase, up to $t \sim 2.3$, 
the effect of the subgrid scales onto the resolved ones 
becomes important, and the LES models correctly reproduce the DNS dynamics again. 
Differences then start to appear between the LES energy approximations 
and the DNS energy, as all scale intensities increase, 
and therefore so does the influence of the intermediate and
highly nonlocal transfer terms (see Section II.B). 
However, their mean values stay close to the DNS energy (see $U_{rms}$ in Table~\ref{table1}).
The same remarks hold for the temporal evolution of the kinetic helicity.
However, in this latter phase, $t \geq 5$, one can note that the LES PH model provides 
a slightly better approximation than the LES P one, for both energy and helicity.
When computing the temporal mean of the relative error between the 
LES and DNS data, we obtain for the energy $1.28\%$ for the helical model 
(versus $1.36\%$ for LES P). For the helicity, these  errors are respectively $2.20\%$
for LES PH and $2.27\%$ for LES P. Considering that the cost of computing the additional 
helical term is rather small (the LES PH needs $6\%$ CPU time more than the LES P simulation),
the slight improvement when using the helical model is worth considering;
in particular, note that it reproduces better the temporal  oscillatory
variation of the total energy, although at a higher intensity. On the
other hand, the fact that the non-helical model performs almost as well
shows that helicity does not play a significant role in the small scales,
in agreement with the statistical argument of return to
isotropy in the small scales, and with the fact that the
relative helicity decays faster than $k^{-1}$ in the small scales
which are being truncated in an LES computation.

\begin{figure}[ht]
   \centerline{\epsfig{file=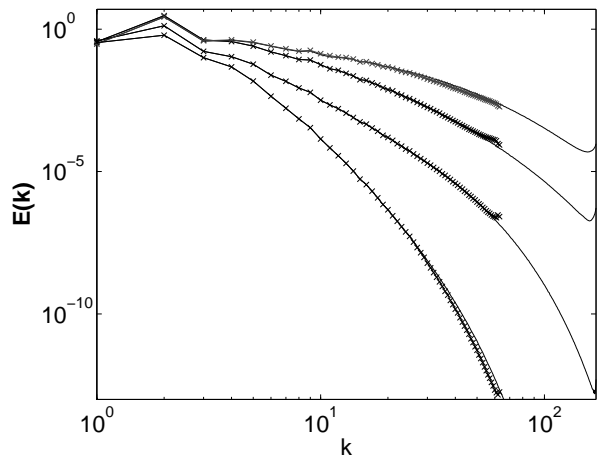,
   width=\linewidth}}
     \caption{Temporal development of energy spectra $E(k,t)$ shown at $t=0.5$, $t=1.0$, $t=2.0$ and $t=3.0$, 
              for runs {\bf IV} (DNS, solid line) and {\bf V} (LES-PH, plusses).}
   \label{Spectre_evolution}
\end{figure}
\noindent

The temporal behavior of the LES flows can be understood when looking at kinetic energy 
spectra at early times, as plotted in Fig.~\ref{Spectre_evolution}.
Instantaneous DNS energy spectra are well fitted by LES spectra up to $t \sim 3.0$, including in the phase of development toward a Kolmogorov spectrum. 
At larger times, the instantaneous modeled slightly spectra diverge from the DNS ones , as seen in the steady state in the previous section, with
small scales slightly underestimated and large scales slightly overestimated.    
\vspace{10pt}
\subsection{Statistical analysis}
\begin{figure}[ht]
\includegraphics[width=7cm, height=49mm]{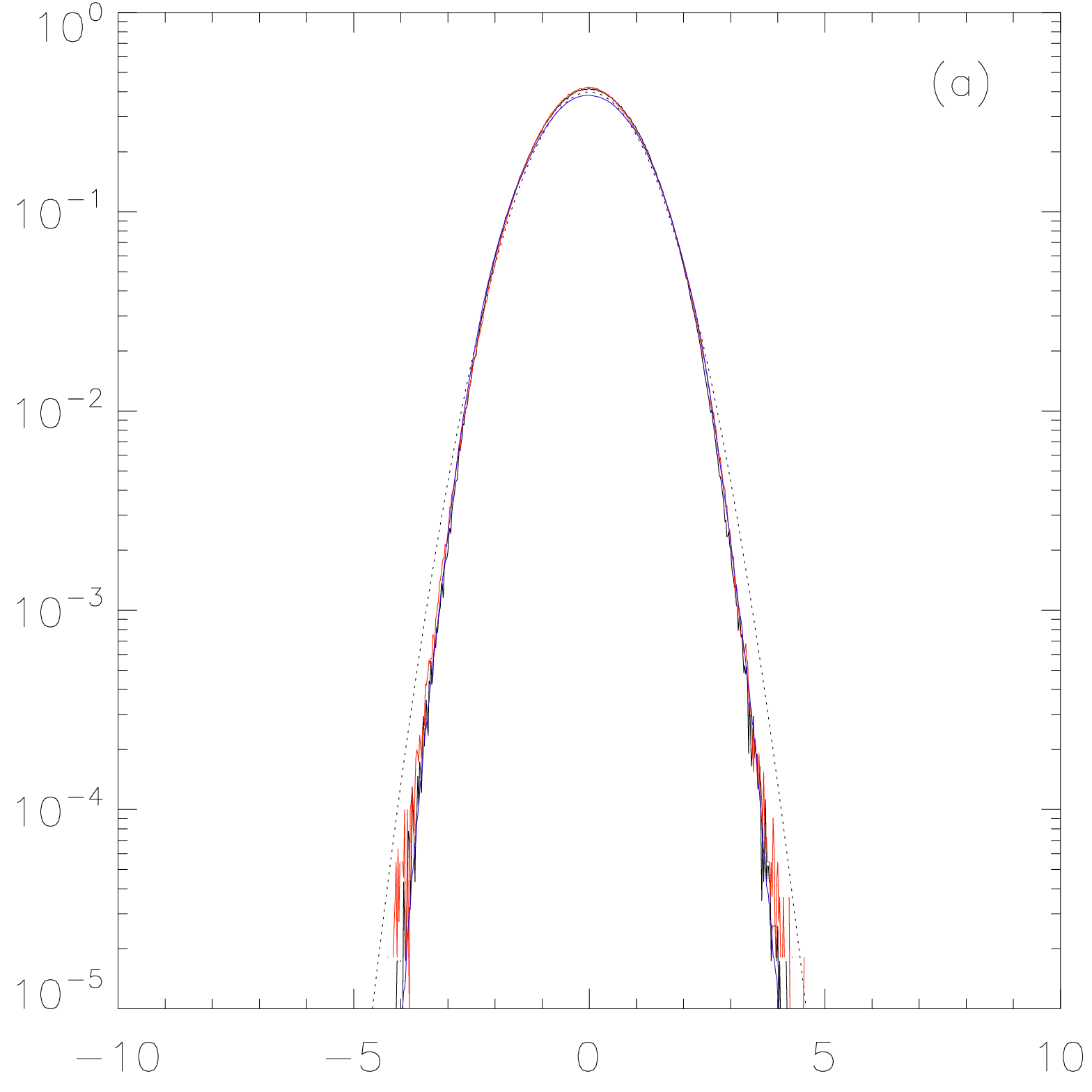}\\
\includegraphics[width=7cm, height=49mm]{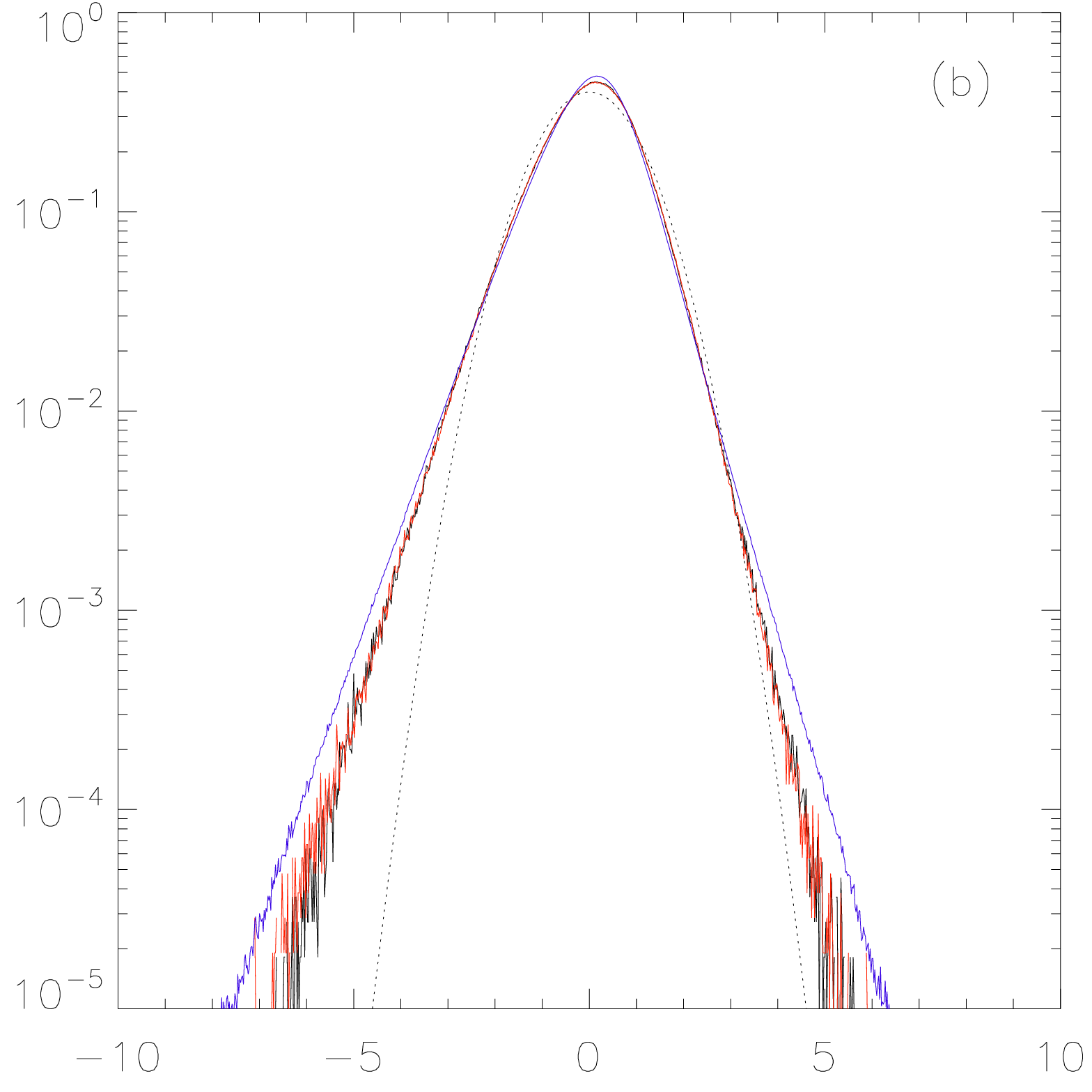}\\
\includegraphics[width=7cm, height=49mm]{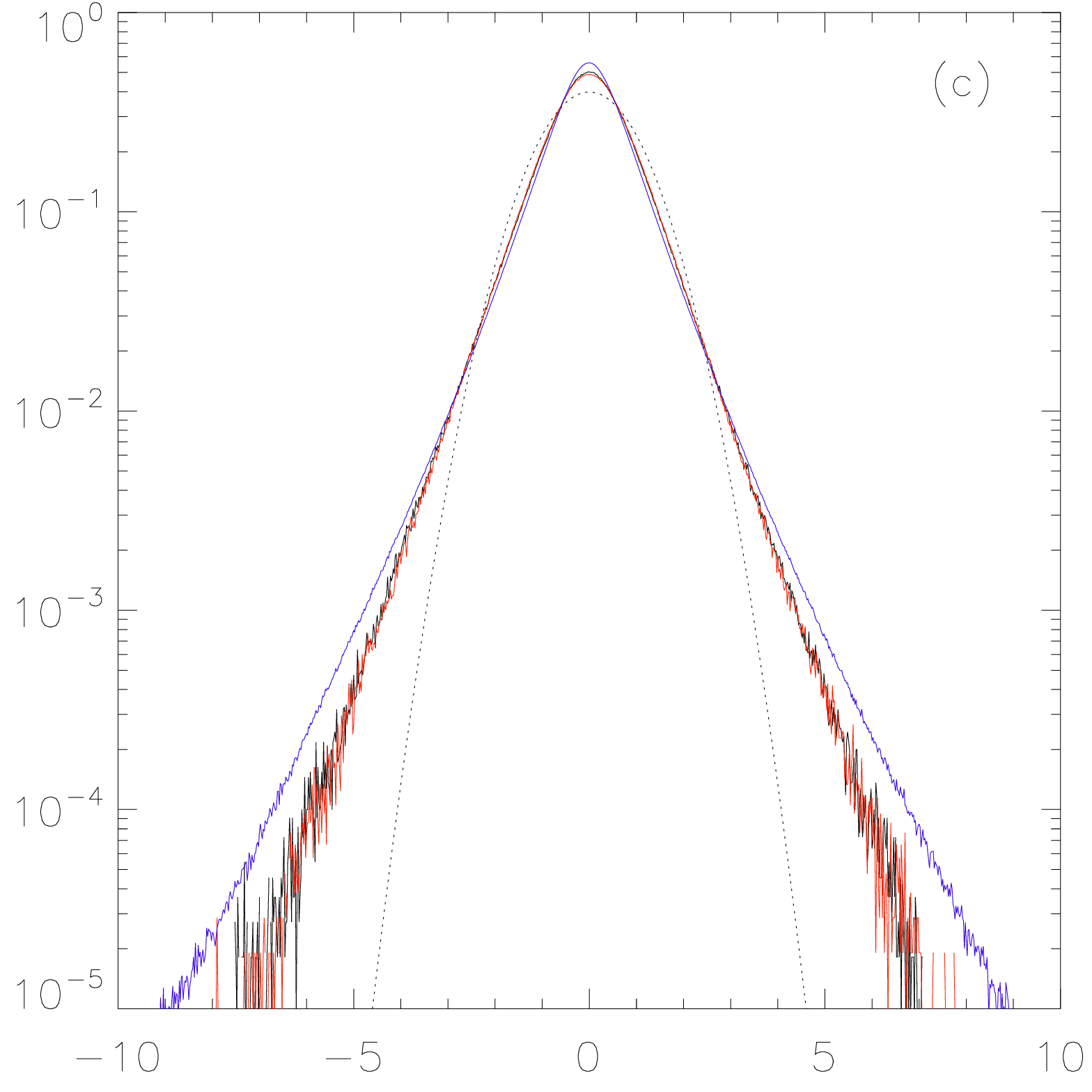}
\caption{Mean probability distributions of $v_x$ (a), ${\partial v_x}/{\partial x}$ (b) and ${\partial v_y}/{\partial x}$ 
(c), normalized so that $\sigma=1$, for 
data {\bf I} ($256^3$ DNS, in blue), {\bf Ir} ($64^3$ reduced DNS data, 
in black) and {\bf II} ($64^3$ LES PH, in red), 
shown together with a Gaussian distribution (dotted line). See Table~\ref{table1}.
}
\label{pdf} 
\end{figure}

Flow statistics are now investigated.
Probability density functions (hereafter, {\it pdf}), are computed from $20$ velocity snapshots extracted each
$3 \tau_{NL}$ and spanning the flow steady states, between $t=10.0$ and $t=60.0$. We compare data sets obtained from runs
{\bf{I}} ($256^3$ DNS), {\bf{Ir}} ($64^3$ reduced DNS data) and  {\bf{II}} ($64^3$ LES PH),
for which we have large velocity samples. 
For clarity purpose, since differences between helical and non helical models are not visible on the {\it pdf}, 
we only represent LES PH results versus DNS ones. 
Fig.~\ref{pdf} (a) displays the statistical distribution of the $v_x$-velocity component, after being normalized so that 
$\sigma^2=<v_x^2>=1$ ($\sigma$ being the standard deviation), together with a Gaussian distribution.
The obtained distributions for all data runs are close to Gaussian, typical of large scale velocity behavior.
The LES models being designed for recovering correctly the large-scale flow, 
the LES {\it pdf} are identical to those of the truncated DNS data set {\bf Ir} 
(where the smallest DNS scales have been filtered out for $|{\bf k}| > k_c$). 
Note that they are also very close to the velocity distribution of the full DNS data set. 

Examples of spatial distributions of longitudinal and lateral velocity derivatives, 
${\partial v_x}/{\partial x}$ and ${\partial v_y}/{\partial x}$ respectively, are shown in Fig.~\ref{pdf} (b) and Fig.~\ref{pdf} (c).
The distributions are closer to an exponential than to a Gaussian. This behavior is even more pronounced
for lateral derivatives with a slight departure from an exponential law. 
The wings of the {\it pdf} are mainly due to small-scale 
velocity gradients. Since the DNS flow has more excited small scales, the wings of the 
associated velocity derivatives distributions
are more extended than for LES data. 
The LES distributions correctly reproduced the DNS ones up to $3 \sigma$,
however there is almost no differences with the {\it pdf} of the {\bf Ir}  data set.  
\begin{table}[h!]
\caption{\label{table2}Temporal mean skewness of velocity derivatives 
for runs {\bf I}-{\bf III} and {\bf Ir} data (see Table~\ref{table1}). 
Error bars are computed from instantaneous data.}
\begin{ruledtabular}
\begin{tabular}{ccccc}
& {\bf I} DNS &{\bf{Ir}} & {\bf{II}} LES PH & {\bf{III}} LES P\\
\hline
${\partial v_x}/{\partial x}$ & $-0.45 \pm 0.05$ & $-0.35 \pm 0.05$ &
$-0.35 \pm 0.04$ &
$-0.33 \pm 0.05 $ \\
${\partial v_y}/{\partial y}$ & $-0.45 \pm 0.06$ & $-0.34 \pm 0.04$ &
$-0.34 \pm 0.06$ &
$-0.34 \pm 0.06 $ \\
${\partial v_z}/{\partial z}$ & $ -0.46 \pm 0.06$ & $-0.35 \pm 0.05$ &
$-0.34 \pm 0.07$ &
$-0.33 \pm 0.05 $ \\
\end{tabular}
\end{ruledtabular}
\end{table}

\begin{table}[h!]
\caption{\label{table3}Temporal mean flatness of  
the velocity gradients for the same runs as in Table~\ref{table2}.}
\begin{ruledtabular}
\begin{tabular}{ccccc}
& {\bf I} DNS &{\bf{Ir}} & {\bf{II}} LES PH & {\bf{III}} LES P\\
\hline
${\partial v_x}/{\partial x}$  & $5.0 \pm 0.2$ & $4.0 \pm 0.2$ & $4.0 \pm 0.3$ &
 $4.0 \pm 0.3$\\
${\partial v_y}/{\partial x}$  & $7.2 \pm 0.6$& $5.0 \pm 0.4$ & $4.9 \pm 0.3$ &
$4.8 \pm 0.4$ \\
${\partial v_z}/{\partial x}$  & $7.3 \pm 0.4$ & $5.1 \pm 0.4$ & $4.9 \pm 0.6$ &
$4.8 \pm 0.5$ \\
${\partial v_x}/{\partial y}$  & $ 7.4 \pm 0.4$ & $5.0 \pm 0.3$ & $4.9 \pm 0.5$ &
 $4.8 \pm 0.5$\\
${\partial v_y}/{\partial y}$  & $5.1 \pm 0.2$ & $3.9 \pm 0.1$ & $3.9 \pm 0.2$ &
$3.9 \pm 0.2$ \\
${\partial v_z}/{\partial y}$  & $7.3 \pm 0.6$& $5.0 \pm 0.3$ & $4.8 \pm 0.4$ &
$4.7 \pm 0.5$ \\
${\partial v_x}/{\partial z}$  & $7.3 \pm 0.6$& $5.0 \pm 0.5$ & $4.9 \pm 0.5$ &
 $4.7 \pm 0.4$\\
${\partial v_y}/{\partial z}$  & $7.3 \pm 0.9$& $5.0 \pm 0.4$ & $4.8 \pm 0.7$ &
$4.8 \pm 0.8$ \\
${\partial v_z}/{\partial z}$  & $5.1 \pm 0.4$& $4.0 \pm 0.2$ & $3.9 \pm 0.3$ &
$3.9 \pm 0.3$ \\
\end{tabular}
\end{ruledtabular}
\end{table}
In order to quantify the distributions of the velocity fluctuations, and their differences 
among DNS, LES PH and LES P data, we compute low-order moments, namely the
skewness ($S_3$) and flatness ($S_4$) factors of the velocity derivatives, 
defined as $S_n=<f^n>/<f^2>^{n/2}$ where $f$ stands for any velocity derivatives.
Their temporal means and error bars, are given in Table~\ref{table2} and Table~\ref{table3} respectively. 
For a fair comparison with the results of our LES simulations, we also compute the skweness and flatness
factors based on the {\bf Ir} reduced DNS velocity fields.
The {\it pdf} of the longitudinal velocity derivatives present an asymmetry that yields their well-known negative 
skweness, with  $S_3 \sim -0.45$ for DNS velocities, a value comparable to other simulations at $\Rla \sim 500$
(e.g. \cite{sreeni}). The skweness for the reduced DNS data (from $256^3$ to $64^3$) is about $22 \%$ lower, a reduction
simply due to the truncation of velocity fields in Fourier space as noted in \cite{dubois}.
The LES models give almost identical results, with in most cases, slightly closer values for LES PH data sets.  
It remains an open problem to know whether this type of agreement persists for higher Reynolds numbers.
For all runs, the lateral velocity derivatives are much more symmetric: $S_3 \sim 0$ (with fluctuations varying from 
$10^{-2}$ to $10^{-4}$), as expected for fields that are almost statiscally isotropic \cite{monin}.
The longitudinal and lateral velocity derivatives do not have the same $S_4$ flatness factors. 
For the DNS data, the flatness values for ${\partial v_x}/{\partial x}$, ${\partial v_y}/{\partial y}$ 
and ${\partial v_z}/{\partial z}$ are close to $5$, while the flatness for the lateral velocity 
derivatives (for example ${\partial v_x}/{\partial y}$ and ${\partial v_x}/{\partial z}$)
are larger, with values around $7$. The reduced {\bf Ir} data present a loss of 
$\sim 20 \%$ and $\sim 30 \%$ for longitudinal and lateral derivatives, respectively. 
Once again, both LES data provide similar values with a better approximation 
of the lateral derivatives for the LES PH computation.

\vspace{10pt}
\subsection{Visualization in physical space}
The topological properties of the different flows are now investigated: 
comparisons between DNS and LES PH computations are carried out on either instantaneous 
and or mean velocity fields. 
\begin{figure}[h!]
\begin{minipage}[t]{.49\linewidth}
    \begin{center}
       \includegraphics[width=4cm]{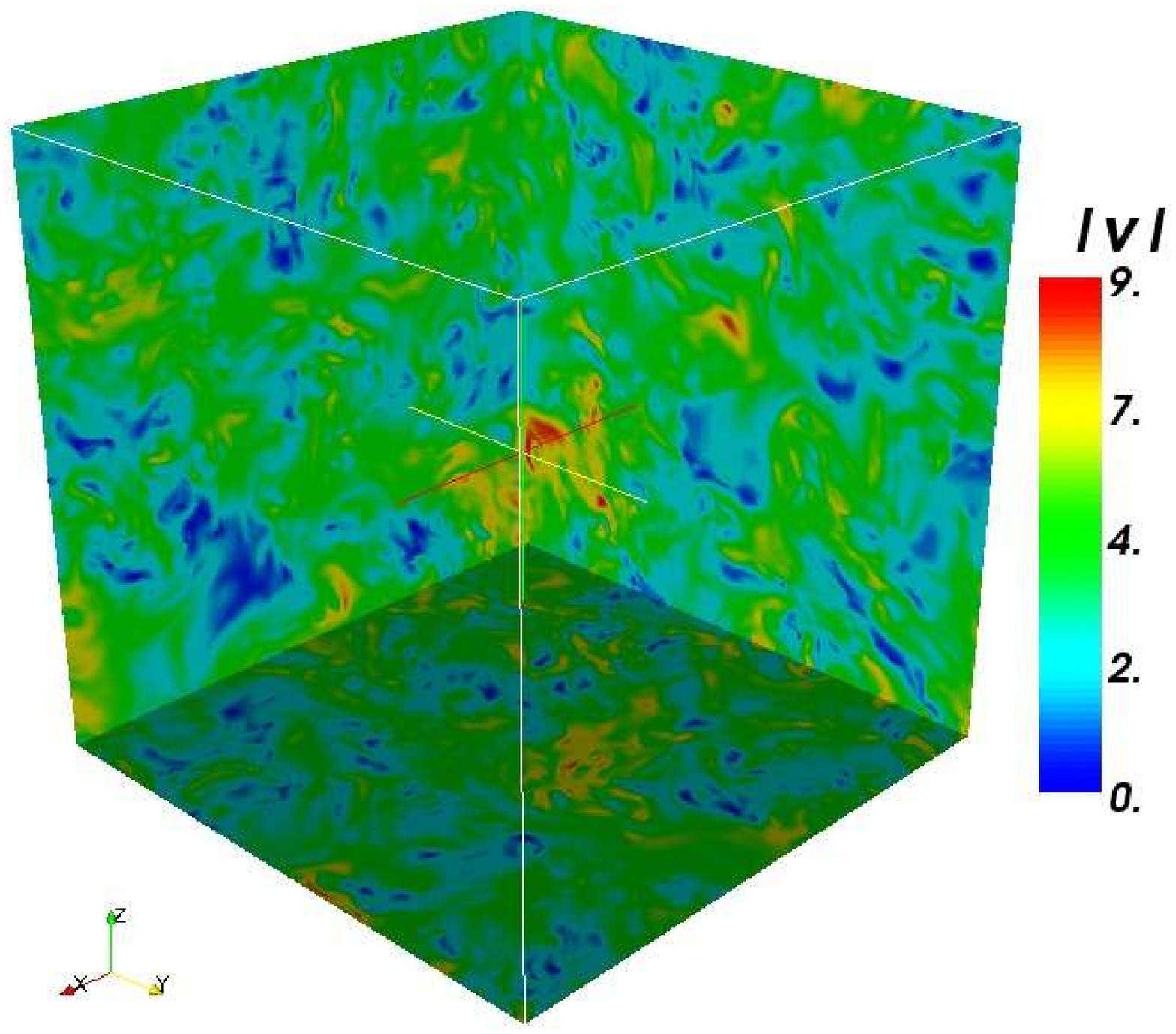}
    \end{center}
\end{minipage}
\hfill
\begin{minipage}[t]{.49\linewidth}
    \begin{center}
       \includegraphics[width=4cm]{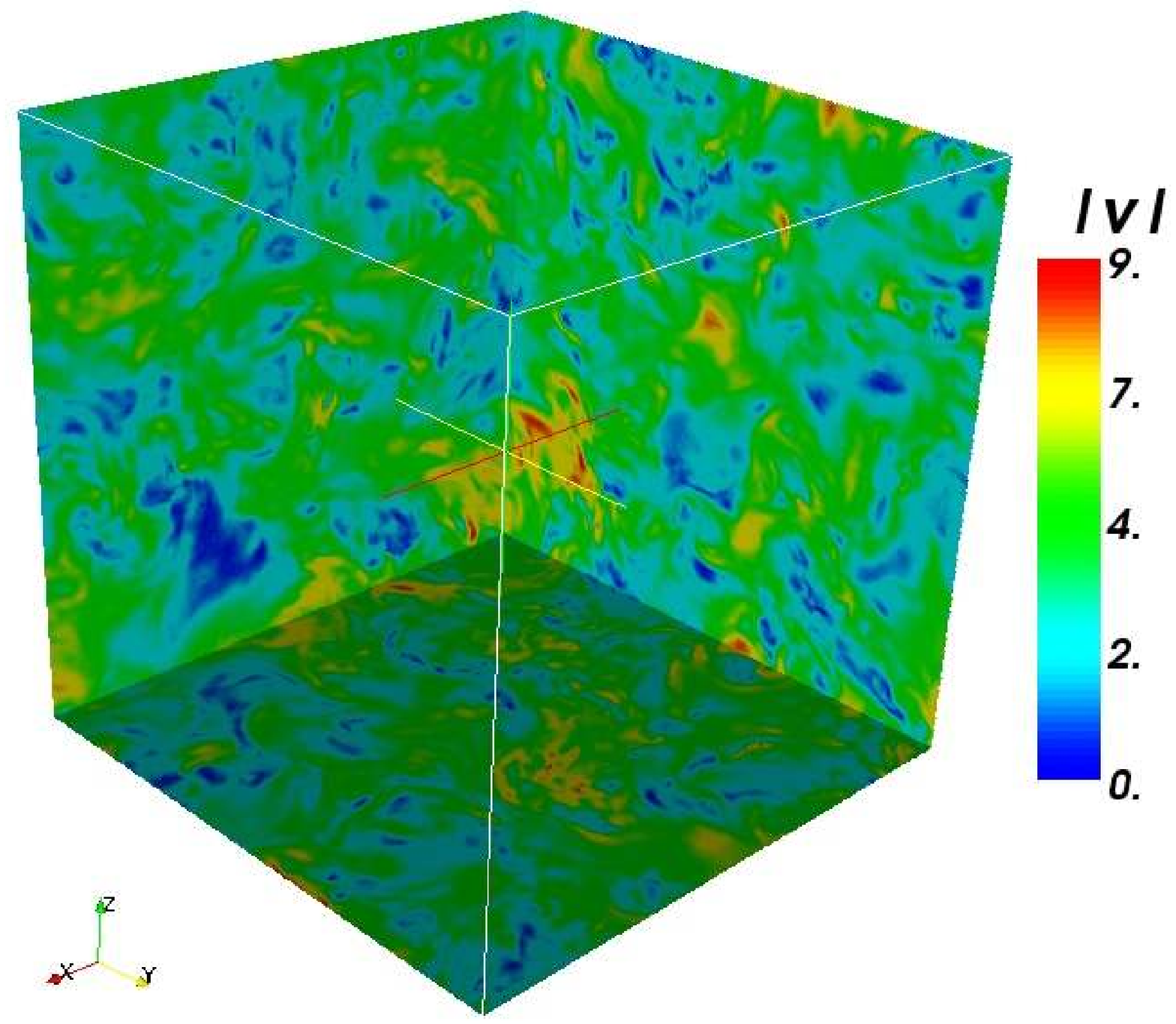}
    \end{center}
\end{minipage}
\caption{
Contour plots of the velocity intensity $\mid {\bf v}({\bf x},t) \mid$ at time $t=2.8$
on the boundaries of the periodic box, for run {\bf IV} ($512^3$ DNS, left) and run {\bf V} ($128^3$ LES PH, right).
}
       \label{v_instantane}
\end{figure}

For runs {\bf IV} ($512^3$ DNS) and {\bf V} ($128^3$ LES PH), Fig.~\ref{v_instantane} 
displays contour plots of the velocity intensity at time $t=2.8$, shown
on three sides of the periodic box. Although our LES model cannot exactly
reproduce the DNS flow, one can notice that the main flow structures, 
and their intensities, are well represented  at times before the statisticaly stationary regime.
More precisely, the mean spatial correlation of the pointwise LES PH velocity field with the DNS one
is $84.37\%$,  while it is $84.32 \%$ in the case with the LES P pointwise velocity at the same time (not shown).
\begin{figure}[h!]
\begin{minipage}[t]{.49\linewidth}
    \begin{center}
       \includegraphics[width=4cm]{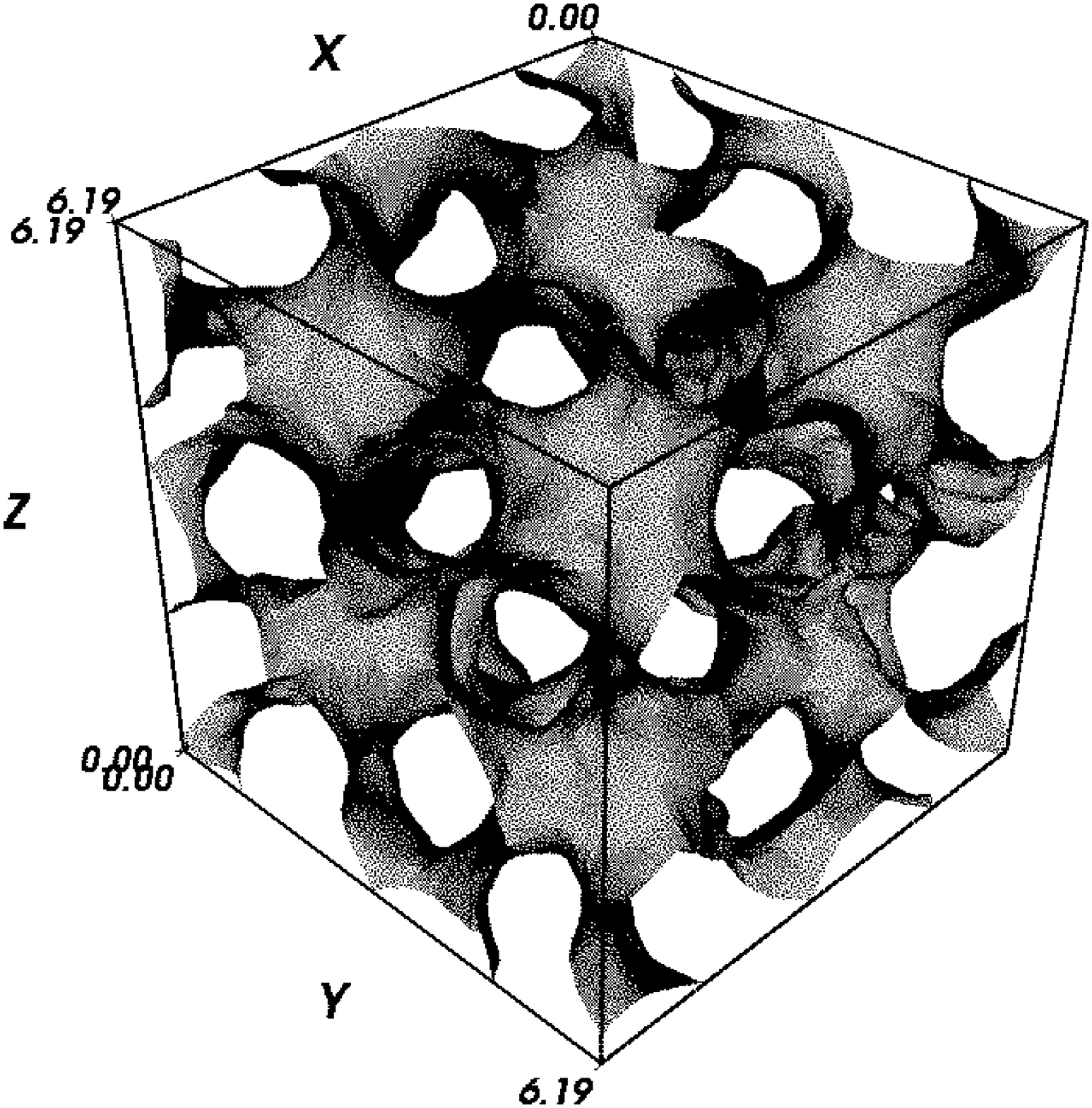}
    \end{center}
\end{minipage}
\hfill
\begin{minipage}[t]{.49\linewidth}
    \begin{center}
       \includegraphics[width=4cm]{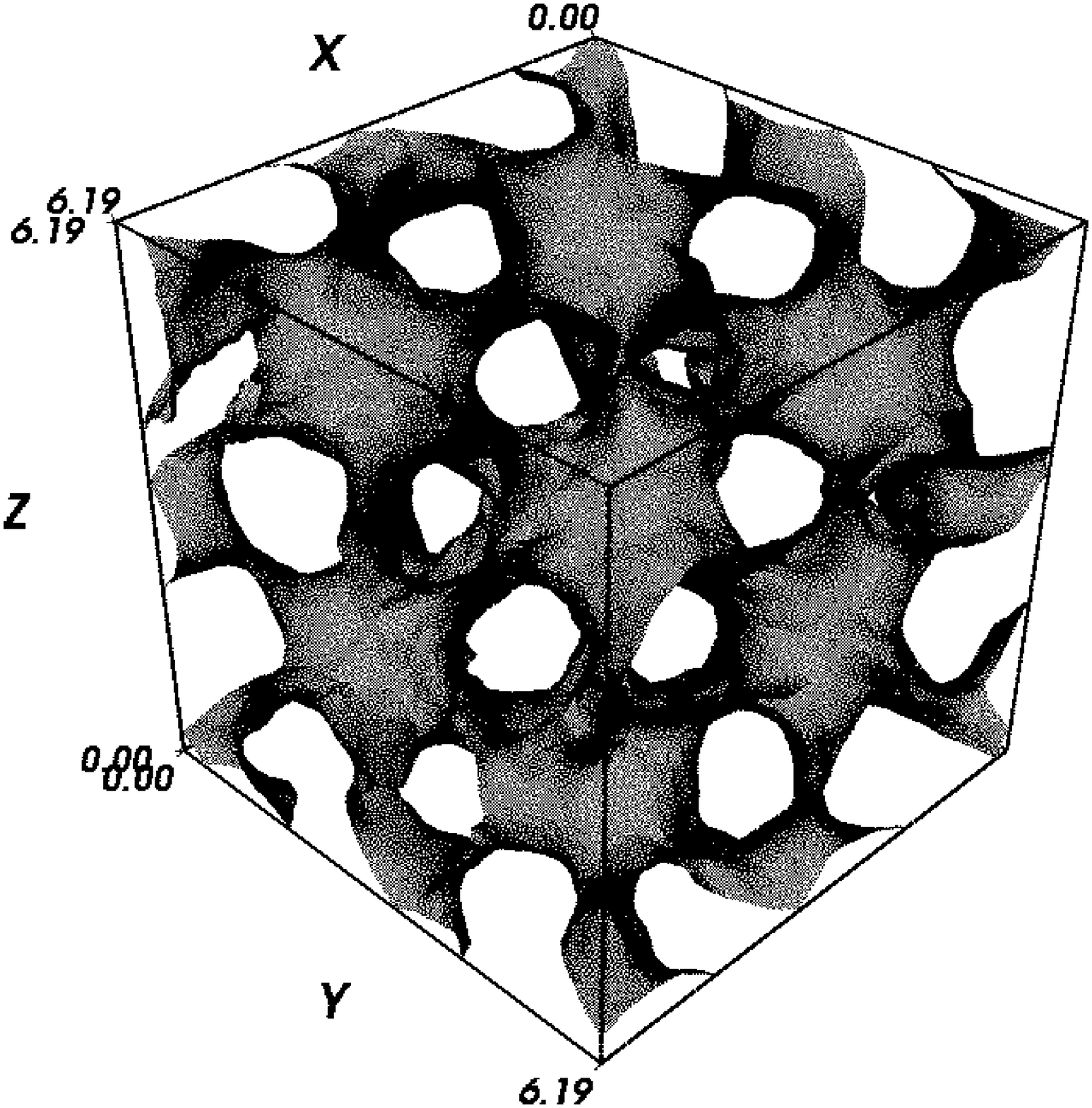}
    \end{center}
\end{minipage}
\caption{Isosurface of the mean flow intensity $\mid <{\bf v}({\bf x},t)> \mid$ for
run {\bf I} ($256^3$ DNS, left) and run {\bf II} ($64^3$ LES PH, right) plotted at a level of $2$, with maximum 
intensity values of $3.18$ for run {\bf I} and $3.31$ for run {\bf II}. Averages are taken over $50$ 
computational times spanning the steady states.
}
       \label{iso_visu}
\end{figure}

For runs {\bf I} ($256^3$ DNS) and {\bf II} ($64^3$ LES PH), Fig. \ref{iso_visu} shows one isosurface of
mean velocity intensities at a level of $2$, i.e. roughly at $63 \%$ of the maximum intensity for DNS data, 
and at $60 \%$ for LES PH data.
The mean velocity fields are time averaged each $3\tau_{NL}$ during the steady states that correspond 
to $67 \tau_{NL}$ for the DNS flow and $68 \tau_{NL}$ for the LES PH run. 
From these runs, with the longest stationary phases, we observe that 
the mean velocity field, and thus the large and intermediate flow scales, are not alterated by the modeling 
of the transfers linked to the subgrid scales.
Indeed, the mean spatial correlation between the pointwise LES and DNS time averaged velocity fields 
is $91.96 \%$ (for completness, it is $92.38 \%$ when the LES P mean field is considered).

Finally, the flow isotropy is estimated by means of coefficients computed as in \cite{curry}. 
For each wavevector ${\bf k}$, an orthonormal reference frame is defined as 
(${\bf k}/|{\bf k}|$, ${\bf e}_1({\bf k})/|{\bf e}_1({\bf k})|$, ${\bf e}_2({\bf k})/|{\bf e}_2({\bf k})|$),
with ${\bf e}_1({\bf k})= {\bf k} \times {\bf z}$ and ${\bf e}_2({\bf k})={\bf k} \times {\bf e}_1({\bf k})$,
where ${\bf z}$ is the vertical unit wavevector. In that frame, since the incompressibility condition, 
say for the velocity field, yields 
${\bf k} \cdot {\bf v}({\bf k}) = 0$, ${\bf v}({\bf k})$ is only determined by its two components 
${\bf v}_1({\bf k})$ and ${\bf v}_2({\bf k})$. The isotropy coefficient is then defined  
as $C^{\bf v}_{iso}=\sqrt{<|{\bf v}_1|^2>/<|{\bf v}_2|^2>}$, with thus a unit value for fully isotropic flows. 
A similar coefficient can be based on the vorticity field, $C^{\bf w}_{iso}$, characterizing the
small scale isotropy.
Isotropy coefficient values are given in Table~\ref{tableiso} for velocity and vorticity fields of the 
flows visualized in (Fig.~\ref{v_instantane}) and (Fig.~\ref{iso_visu}). 
Instantaneous values are computed at $t=2.8$ for data {\bf IV} ($512^3$ DNS) and {\bf V} ($128^3$ LES PH)  
(Fig.~\ref{v_instantane}), while for the mean flow shown in (Fig.~\ref{iso_visu}), the coefficients are 
based on the time averaged fields of runs {\bf I} ($256^3$ DNS) and {\bf II} ($64^3$ LES PH). 
For comparison, the isotropy coefficients are also computed with the data of the LES P runs 
{\bf VI} ($256^3$) and {\bf III} ($64^3$).
One can see that, altogether, the isotropic properties of the flow are correctly restaured by the LES models. 
\begin{table}[h!]
\caption{\label{tableiso} Isotropy coefficients of velocity, $C^{\bf v}_{iso}$, and vorticity
fields, $C^{\bf w}_{iso}$, for flows shown in (Fig.~\ref{v_instantane}) and (Fig.~\ref{iso_visu}).
For runs {\bf I} to {\bf III}, values are computed from the mean flows.
For runs {\bf IV} to {\bf VI}, they are given at $t=2.8$.
For completeness, the coefficient values are also given for LES P runs. }
\begin{ruledtabular}
\begin{tabular}{cccc}
& & $C^{\bf v}_{iso}$ & $C^{\bf w}_{iso}$ \\
\hline
{\bf I} &DNS & 0.990 & 1.010 \\
{\bf II} &LES PH & 1.026 & 0.979 \\
{\bf III} &LES P & 0.988 & 1.009  \\
{\bf IV} &DNS & 0.961 & 1.033 \\
{\bf V}  &LES PH & 0.955 & 1.035 \\
{\bf VI} &LES P & 0.955 & 1.029 \\
\end{tabular}
\end{ruledtabular}
\end{table}
\vspace{10pt}
\subsection{Comparison with a Chollet-Lesieur approach}
It may be instructive to test our model against another spectral LES approach, also based on the EDQNM closure; we have thus
performed a simulation using a Chollet-Lesieur scheme \cite{chollet_lesieur}
(run {\bf VII} LES CL in Table~\ref{table1}).
The CL model allows energy tranfer from subgrid to resolved scales through a dissipation mechanism, with the
help of a dynamical eddy viscosity $\nu_{CL}(k,t)$ defined as : 
\begin{equation}
\nu_{CL}(k,t)=C\nu^+(k,t)\sqrt{E(k_{cut},t)/k_{cut}} \ ;
\end{equation}  
$k_{cut}=N/2-3$ is the cut-off wavenumber, $N$ being the number of grid points per direction,
and $\nu^+(k,t)$ is the so-called cusp function evaluated as $\nu^+(k,t)=(1+3.58(k/k_{cut})^8)$. 
We recall that $C\sqrt{E(k_{cut},t)/k_{cut}}$ is the asymptotic expression of the nonlocal
tranfers from subgrid to resolved scales, and it assumes a $k^{-5/3}$ Kolmogorov spectrum extending 
to infinity. The constant $C$ is adjusted 
with the Kolmogorov constant computed from the ABC flow resolved by the DNS run using $512^3$ grid points, namely
$C=0.14$.
To be able to compare our DNS and LES simulations, runs {\bf IV} to  {\bf VI} (see Table~\ref{table1}),  to a CL 
simulation with $128^3$ grid points, the kinematic viscosity used for the former runs, $\nu=2.e^{-3}$, is added to 
$\nu_{CL}(k,t)$. The asymptotic value of the eddy viscosity can be obtained as the temporal mean of $\nu_{CL}(0,t)$ 
in the time interval $[3.5,7.0]$ and is here estimated to be $\sim 6.e^{-4}$.
\begin{figure}[ht]
   \centerline{\epsfig{file=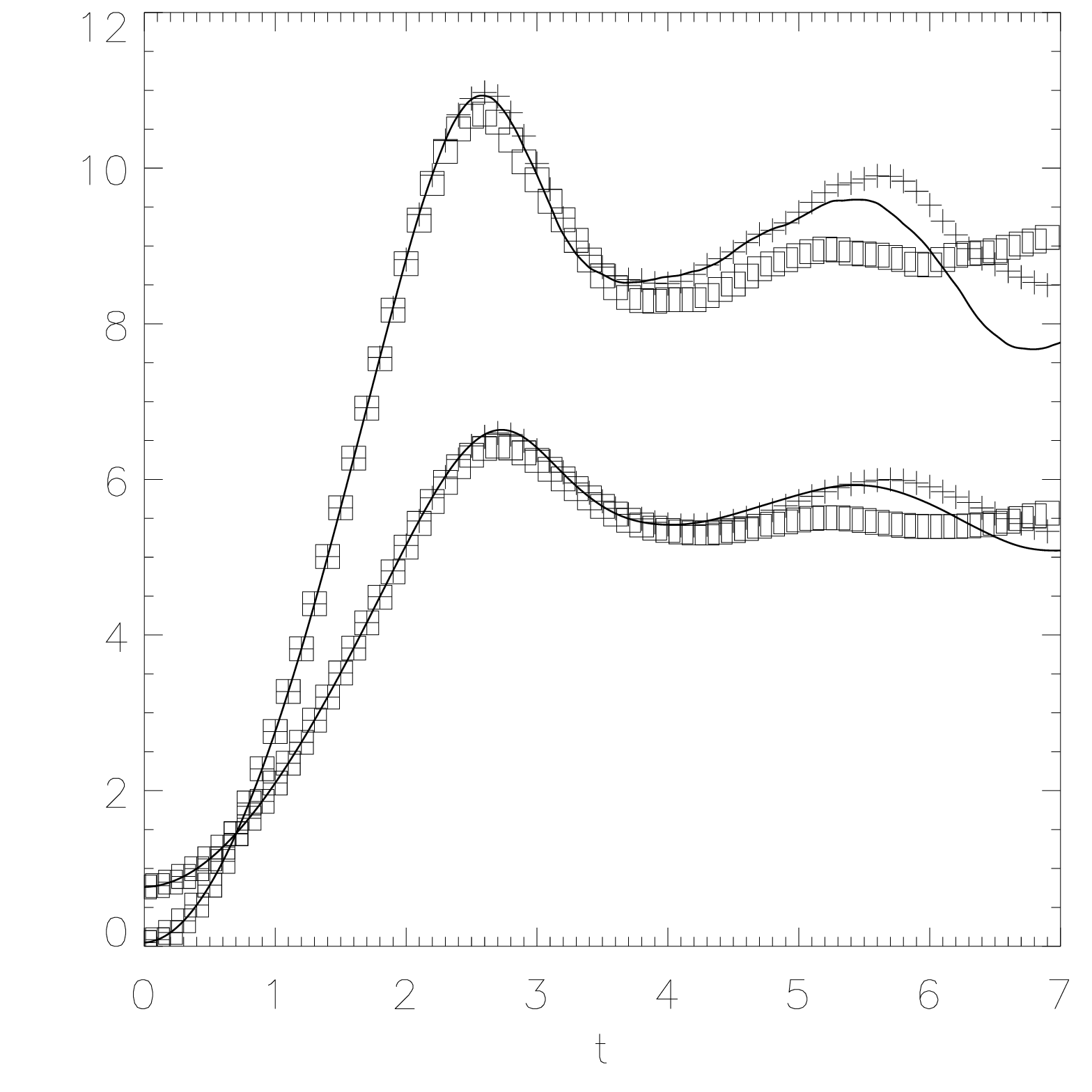, 
    width=\linewidth,height=50mm}}
     \caption{Evolution of energy $E(t)$, lower curves, and helicity $H(t)$, upper curves,
 for data {\bf IV} ($512^3$ DNS, solid line), {\bf V} 
($128^3$ LES PH, plusses) and {\bf VII} ($128^3$ LES CL, squares). }
   \label{compa_evo_DNS_CL_PH}
\end{figure}

Both CL and LES PH runs provide a close agreement with DNS temporal evolutions  of kinetic energy and helicity 
during the growth phase, {\it i.e} up to $t \sim 2.3$ (see Fig.~\ref{compa_evo_DNS_CL_PH}). 
In steady states, although noticeable deviations occur between 
DNS and LES approximations, a slightly better assessment is visible for the LES PH model;
indeed, the peak of oscillation at $t\sim 5.4$ is weaker in run {\bf VII}.
Note that comparisons of LES PH versus LES P data, runs {\bf V} and {\bf VI} respectively,
are already presented in Section III.C.
\begin{figure}[ht]
   \centerline{\epsfig{file=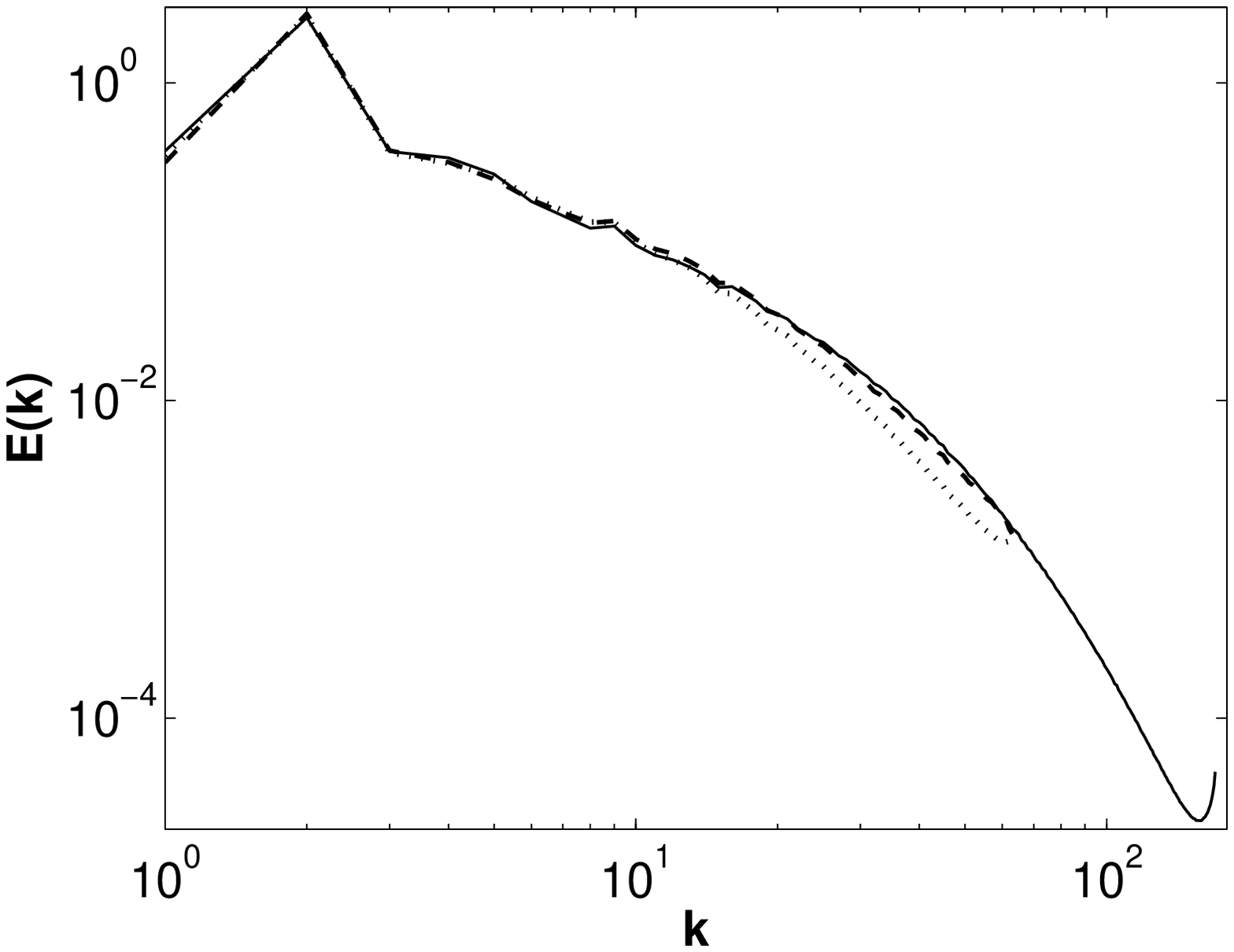, 
    width=\linewidth,height=40mm}}
   \centerline{\epsfig{file=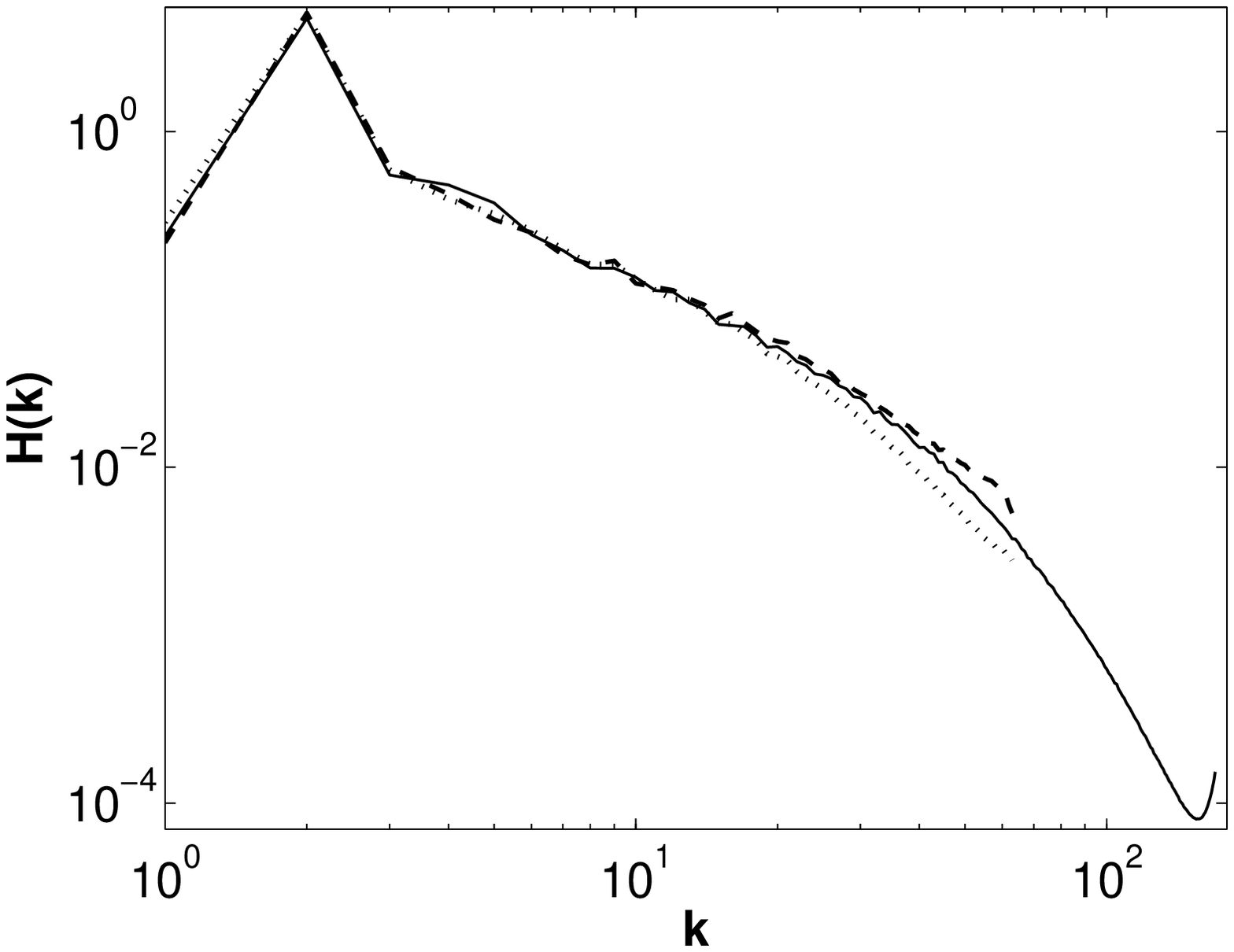, 
    width=\linewidth,height=40mm}}
     \caption{Mean energy (top) and helicity (bottom) spectra for runs {\bf IV} ($512^3$ DNS, solid line), 
{\bf V} ($128^3$ LES PH, dashed line) and {\bf VII} ($128^3$ LES CL, dotted line).
      }
   \label{compa_DNS_CL_PH}
\end{figure}

Mean energy and helicity spectra, time averaged from $t=3.5$ to $t=7.0$, 
are plotted in Fig.~\ref{compa_DNS_CL_PH} for runs {\bf IV}, {\bf V} and {\bf VII} described in Table~\ref{table1}. 
Energy spectra only exhibit a short $k^{-5/3}$ inertial range for all runs, including the CL run, due to the
procedure we chose avoiding a non-zero asymptotic viscosity. 
With  $\nu=0$ in the CL scheme, the results might be different.
However, for both energy and helicity spectra, LES PH data give slightly closer results 
when compared to DNS data, as our CL simulation seems to overestimate positive energy 
transfer from resolved scales (between $k\sim 15$ to $k_{cut}=61$) to subgrid scales. 
\subsection{Predictions for high Reynolds number flow}
In this last section, we present model computations for flows at high Reynolds number. 
Recently, Kurien {\it et al.}  \cite{kurien} 
showed that for flows with maximum helicity, both energy an helicity spectra exhibit a $k^{-4/3}$ scaling range 
following the $k^{-5/3}$ Kolmogorov range and preceding the dissipation range, a result also found in \cite{alex_long}.
This change in the energy spectrum is estimated from energy flux based on a characteristic time scale, denoted ${\tau_H}$,
of distortion (or shear) of eddies with wavenumber $k$ submitted to out-of-plane velocity correlations, 
corresponding to helicity transfer. 
We recall that the two dynamical times in competition are estimated by $\tau_H^2(k) \sim (|H(k)|k^2/2)^{-1}$ and
$\tau_{NL}^2(k) \sim (E(k)k^3)^{-1}$.
Moreover, from DNS of the forced Navier-Stokes equation, these authors associate the well-known bottleneck effect 
with this scaling change when $\tau_H$ becomes physically relevant. 
The ABC flow being known for the presence of strong helical structures,  
we performed two simulations using our LES PH and LES P models at kinematic viscosities $\nu=5.e^{-4}$, 
with $256^3$ grid points (respectively run {\bf VIII} and run {\bf IX} in Table \ref{table1}).
For these flows, the total helicity $H(t)=1/2<{\bf v(t)} \cdot {\bf w(t)}>$, averaged 
over $\sim 8 \tau_{NL}$ in the steady state, is equal to $9.26$ for data set {\bf VIII} and 
to $8.65$ for run {\bf IX}. 
Temporal means of the total relative helicity $H(t)/E(t)$ are also close (within $1 \%$) for both computations, 
namely $0.195$ for the LES PH run versus $0.186$ for the LES P one. 
\begin{figure}[h!]
   \centerline{\epsfig{file=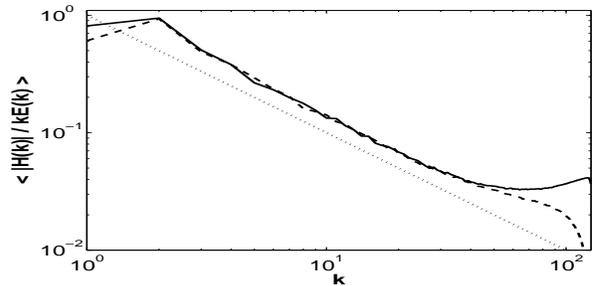,
   width=\linewidth,height=40mm}}
     \caption{Time averaged relative helicity $<|H(k)|/kE(k)>$ for data sets :
{\bf XIII} ($256^3$ LES PH), solid line,  and {\bf IX} ($256^3$ LES P), dashed line. 
A  $k^{-1}$ slope is plotted for comparison (dotted line).
 }
   \label{HR_k}
\end{figure}
More precisely, in the range  $10<k<100$, the relative helicity $|H(k)|/kE(k)$, viewed as
an estimation of the ratio $\tau_H^2(k)/\tau_{NL}^2(k)$,
lies in between  $13.5 \%$ and $3.5 \%$ for run {\bf VIII}, and falls from about $14 \%$ to $2 \%$ 
for run  {\bf IX} (see Fig.~\ref{HR_k}), a typical ratio for strong helical flows \cite{kurien}. 
Note also that the relative helicity obtained by both model scales closely to 
a $k^{-1}$ power law observed in previous DNS experiments \cite{alex_long}.

\begin{figure}[h!]
   \centerline{\epsfig{file=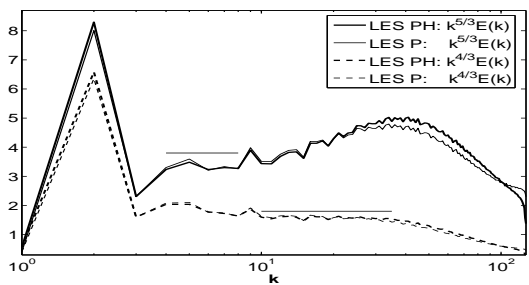,
   width=\linewidth,height=40mm}}
     \caption{Temporal mean of energy spectra compensated by $k^{5/3}$ (solid lines) and 
$k^{4/3}$ (dash lines) for runs {\bf XIII} ($256^3$ LES PH) and {\bf IX} ($256^3$ LES P) (see insert). 
Horizontal segments indicate ranges of $k^{-5/3}$ and  
$k^{-4/3}$ scaling regime. 
 }
   \label{compa_spectreE_p_ph}
\end{figure}

Fig.~\ref{compa_spectreE_p_ph} displays mean energy spectra for the two modeled flows compensated 
by $k^{5/3}$ and $k^{4/3}$ respectively. A $k^{-5/3}$ scaling appears in the range
$4<k<10$, followed by a $k^{-4/3}$ regime in for $10<k<40$, and with no appearance of 
a bottleneck effect, although, in the latter wavenumber interval the estimated 
time ratio $\tau_H/\tau_{NL}$ ranges from $37 \%$ to $20 \%$.
\begin{figure}[h!]
   \centerline{\epsfig{file=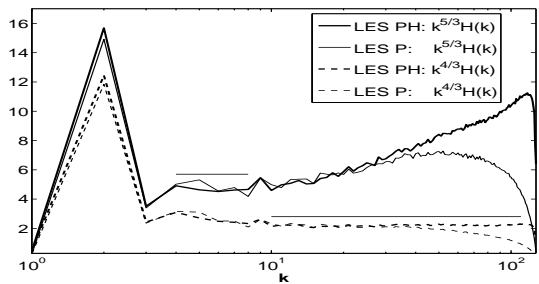,
   width=\linewidth,height=40mm}}
    \caption{Temporal mean helicity spectra compensated by $k^{5/3}$ 
     and $k^{4/3}$ for the same runs as in Fig.~\ref{compa_spectreE_p_ph}. 
}
   \label{compa_spectreH_p_ph}
\end{figure}
Similarly, time averaged helicity spectra compensated by $k^{5/3}$ and $k^{4/3}$ are plotted 
in Fig.~\ref{compa_spectreH_p_ph}.
A $k^{-5/3}$ behavior is seen in approximatively the same range than for energy spectra, while the $k^{-4/3}$ 
scaling occurs from $k \sim 10$ to 
$k \sim 60$ for the LES P flow and up to $k_{c}$, the maximun computational wavenumber, for the LES PH flow. 
Recall that, at high wavenumbers, our LES PH model slightly overestimates helicity spectra 
while our LES P model underestimates them.
However, both LES models well reproduce the observed spectral behaviors obtained from DNS of
flows at lower kinematic viscosities ($\nu=1.e^{-4}$ and $\nu=0.35e^{-4}$) in \cite{kurien}.

\vspace{15pt}
\section{Conclusion}

In this paper, we derive a consistent numerical method, based on the EDQNM closure, 
to model energy interactions between large and small scales for the Navier-Stokes equation. 
As no spectral behavior is {\it a priori} given, our dynamical LES method allows for the modeling 
various flows, whether turbulent or not, compared to former spectral models which can, in principle, simulate 
only infinite Reynolds number flows.
The phase relationships of the small scales are taken into account through a numerical
reconstruction scheme for the spectral velocity field.
Helical effects in turbulents flows, like in vortex filaments, are also considered 
through the evaluation of the energy and helicity transfers.
For this purpose, an "helical eddy diffusivity", similar to an eddy viscosity, and the emission transfer
terms in which the helicity spectrum appears, are incorporated in a second model.    
Numerical tests of our two methods, with and without helical effects included, are performed against 
DNS computations. The spectral, statistical and spatial behaviors at large and intermediate scales
of DNS flows are well restaured in both modeled flows.
We notice some advantages for the model including helical effects, in particular, concerning 
the evaluation of the helicity spectra and the probability distributions of the lateral velocity
field gradients. LES PH also predicts a 4/3 spectrum for the helicity all the way to the cut-off, a point that will need further study.
However, in our approach, we need to calculate at each time step the non local energy (and helicity) 
transfers, with an increased  computational 
cost (roughly by a factor 2), when
compared with other spectral LES models \cite{Chasnov,chollet_lesieur}.  
Whereas the role of helicity in fluids is not necessarily dynamically dominant, such is not the case 
in MHD where both kinetic and magnetic helicity play a prominent role, the former in the dynamo 
process and the latter in its undergoing an inverse cascade to large scales; furthermore, in MHD, 
spectra are not necessarily Kolmogorovian and the models presented in this paper may 
be of some use in MHD as well. 
The extension to MHD turbulent flows, with coupled velocity and magnetic fields, presents
no major difficulties and is under study.

\begin{acknowledgments}
We thank P.D. Mininni for useful discussions.
This work is supported by INSU/PNST and PCMI Programs and CNRS/GdR Dynamo.
Computation time was provided by IDRIS (CNRS) Grant No. 070597, 
and SIGAMM mesocenter (OCA/University Nice-Sophia Antipolis).
\end{acknowledgments}

\appendix
\section{Closure expressions of transfer terms}
For completeness, we recall here the expressions of the nonlinear transfer terms for the energy and the helicity,
$S_E(k,p,q,t)$ and $S_H(k,p,q,t)$ respectively, under the EDQNM closure assumption \cite{PFL} : 
\begin{eqnarray}
\widehat T_{E}(k,t)& = &
\iint_{\Delta}\theta_{_{kpq}}(t)S_E(k,p,q,t)dpdq  \ ,\label{teedqnm}\\
\widehat T_{H}(k,t)& = & 
\iint_{\Delta}\theta_{_{kpq}}(t)S_H(k,p,q,t)dpdq  \ . \label{thedqnm}
\end{eqnarray}
where $\Delta$ is the integration domain with $p$ and $q$ such that ($k,p,q$) form a triangle, 
and $\theta_{_{kpq}}(t)$ is the relaxation time of the triple velocity correlations. 
As usual \cite{lesieur_book}, $\theta_{_{kpq}}(t)$ is defined as :
\begin{equation}
\theta_{_{kpq}}(t)=\frac{1-e^{-(\mu_k+\mu_q+\mu_p)t}}{\mu_k+\mu_q+\mu_p} \ ,
\end{equation}
where $\mu_k$ expresses the rate at which the triple correlations evolve, 
i.e. under viscous dissipation and nonlinear shear. It can be written as:
\begin{equation}
\mu_k=\nu k^2 + \lambda\Big(\int_0^k q^2 E(q,t) dq\Big)^{1/2} \ .
\end{equation}
Here $\lambda$ is the only open parameter of the problem, taken equal to $0.36$ to recover 
the Kolmogorov constant $C_K=1.4$. 
The expressions of $S_E(k,p,q,t)$ and $S_H(k,p,q,t)$ can be further explicited 
(with the time dependency of energy and helicity spectra omitted here) as :
\begin{eqnarray}
S_E(k,p,q,t) & = &\frac{k}{pq}b\left[k^2E(q)E(p)-p^2E(q)E(k)\right]\nonumber \\
& - & \frac{k}{p^3q}c\left[k^2H(q)H(p)-p^2H(q)H(k)\right] \nonumber \\
& = & S_{E_1}(k,p,q,t) + S_{E_2}(k,p,q,t)\nonumber \\
& + & S_{E_3}(k,p,q,t) + S_{E_4}(k,p,q,t)  \ .
\label{S_E}
\end{eqnarray}
Here, $S_{E_1}(k,p,q,t)$, $S_{E_2}(k,p,q,t)$, $S_{E_3}(k,p,q,t)$
and $S_{E_4}(k,p,q,t)$ are respectively used to denote the four terms of the extensive
expression of $S_E(k,p,q,t)$. Note that $S_{E_3}(k,p,q,t)$ and $S_{E_4}(k,p,q,t)$ are absent in the fully 
isotropic case (without helicity), and, of course,  all $S_{H_i}(k,p,q,t)$ terms below.
\begin{eqnarray}
S_H(k,p,q,t)  & = & \frac{k}{pq}b\left[k^2E(q)H(p)- p^2E(q)H(k)\right]\nonumber \\
& - & \frac{k^3}{pq}c\left[E(p)H(q)- H(q)E(k\right)]\nonumber \\
& = & S_{H_1}(k,p,q,t) + S_{H_2}(k,p,q,t)\nonumber \\
& + & S_{H_3}(k,p,q,t) + S_{H_4}(k,p,q,t), 
\label{S_H}
\end{eqnarray}
with analogous short notations as before.

In Eqs.~(\ref{S_E}) and (\ref{S_H}), the geometric coefficients $b(k,p,q)$ and $c(k,p,q)$ 
(in short, $b$ and $c$) are defined as:
\begin{equation}
b=\frac{p}{k}(xy+z^3),\quad c=\frac{p}{k}z(1-y^2) \ ,
\end{equation}
where $x$, $y$, $z$ are the cosines of the interior angles opposite to ${\bf k},{\bf p},{\bf q}$.

Let us now introduce a cut-off wavenumber $k_c$, and define the following three zones for the
integration domain $\Delta$ of Eqs.~(\ref{teedqnm}) and (\ref{thedqnm}):
the inner zone $\Delta^<$ (with $k$, $p$, and $q$ all smaller than $k_c$, the cut-off wavenumber), 
the buffer zone $\Delta^>$ (with $p$ and/or $q$ between $k_c$ and $3k_c$), and 
the outer zone $\Delta^{>>}$ (with $p$ and/or $q$ larger than  $3k_c$); then, 
the boundaries of the transfer term integrals have to be adapted.

In the inner zone corresponding to the fully resolved flow, the resolved transfers write:
\begin{widetext}
\begin{eqnarray}
\widehat{T}_{E}^<(k,t) & = &
\int_{|{\bf k}|\leq k}\!\!\!\!\!\!\!\!-iP_{\alpha \beta}(\textbf{k})k_\gamma \int_{\bf 0}^{|{\bf k}|\leq k_c}\!\!\!\!\!\!\!\! 
v_{\beta}(\textbf{p},t)v_{\gamma}(\textbf{k}-\textbf{p},t)v_{\alpha}(\textbf{-k},t) d\textbf{p} d\textbf{k} \\
\widehat{T}_{H}^<(k,t) & = &
\int_{|{\bf k}|\leq k}\!\!\!\!\!\!\varepsilon_{\alpha \delta \beta}k_\delta k_\gamma 
\int_{\bf 0}^{|{\bf k}|\leq k_c}\!\!\!\!\!\! 
v_{\beta}(\textbf{p},t)v_{\gamma}(\textbf{k}-\textbf{p},t)v_{\alpha}(\textbf{-k},t) d\textbf{p} d\textbf{k} 
\end{eqnarray}
\end{widetext}
where $P_{\alpha \beta}(\textbf{k})=\delta_{\alpha\beta}-k_\alpha k_\beta/k^2$ is the projector on solenoidal vectors, as stated before.

The transfers of energy and helicity between the buffer zone and the inner zone
become: 
\begin{eqnarray}
\widehat{T}_{E}^>(k,t) & = &
\int_{k_c}^{3k_c}\!\!\!\int_{k-p}^{k+p}\!\!\! \theta_{_{kpq}}(t)S_{E}(k,p,q,t)dpdq \quad  \label{teebuffer} \\
\widehat{T}_{H}^>(k,t) & = &
\int_{k_c}^{3k_c}\!\!\!\int_{k-p}^{k+p}\!\!\!\theta_{_{kpq}}(t)S_{H}(k,p,q,t)dpdq    \quad \label{thebuffer}
\end{eqnarray} 
and the transfers of energy and helicity between the outer zone and the inner zone
read:
\begin{eqnarray}
\widehat{T}_{E}^{>>}(k,t) &=&
\int_{3k_c}^{\infty}\!\!\!\int_{k-p}^{k+p}\!\!\! \theta_{_{kpq}}(t)S_{E}(k,p,q,t)dpdq \quad \label{teeout} \\
\widehat{T}_{H}^{>>}(k,t) &=&
\int_{3k_c}^{\infty}\!\!\!\int_{k-p}^{k+p}\!\!\!\theta_{_{kpq}}(t)S_{H}(k,p,q,t)dpdq  \quad \label{theout} 
\end{eqnarray} 

\section{Numerical implementation of the model}
As a first step, the Navier-Stokes equation is solved with the eddy viscosity and the helical 
eddy diffusivity $\nu(k|k_c,t)$ and $\widetilde \nu(k|k_c,t)$ (see Eq.~(\ref{NSfmodel}). 
At this intermediate stage, we obtain a partial estimation of the time updated velocity 
field, since the emission transfer term are not yet taken in to account. 
From this intermediate velocity, we compute the corresponding energy and helicity density fields, 
say at wavevector ${\bf k}$. 
They are then corrected with the appropriate EDQNM emission terms 
according to Eqs.~(\ref{be-system}) and (\ref{bh-system}). 
In the time advance, the next step consists in reconstructing the three 
velocity components from these updated energy and helicity,
$\mathcal{E}(\textbf{k},t)$ and $\mathcal{H}(\textbf{k},t)$ respectively. 
When the velocity components are expressed as 
$v_\alpha(\textbf{k},t)=\rho_\alpha(\textbf{k},t) e^{i\phi_\alpha(\textbf{k},t)}$, 
the incompressibility condition leads to the following system of equations for the
velocity phases and amplitudes (in short $\phi_{\alpha}$ and $\rho_{\alpha}$):
\begin{eqnarray}
\rho_2 \rho_3 cos(\phi_{23})&=&\frac{k_1^2 2\mathcal{E}(\textbf{k})
 - \rho_2^2(k_1^2+k_2^2) - \rho_3^2(k_1^2+k_3^2)}{2k_2k_3} \nonumber \\
\rho_1 \rho_3 cos(\phi_{31})&=&\frac{k_2^2 2\mathcal{E}(\textbf{k})
 - \rho_1^2(k_1^2+k_2^2) - \rho_3^2(k_2^2+k_3^2)}{2k_1k_3}\nonumber\\
\rho_1 \rho_2 cos(\phi_{12})&=&\frac{k_3^2 2\mathcal{E}(\textbf{k})
 - \rho_1^2(k_1^2+k_3^2) - \rho_2^2(k_2^2+k_3^2)}{2k_1k_2}\nonumber\\
\rho_2 \rho_3 sin(\phi_{23})&=&\frac{k_1^2\mathcal{H}(\textbf{k})}{k^2}\nonumber\\
\rho_1 \rho_3 sin(\phi_{31})&=&\frac{k_2^2\mathcal{H}(\textbf{k})}{k^2}\nonumber\\
\rho_1 \rho_2 sin(\phi_{12})&=&\frac{k_3^2\mathcal{H}(\textbf{k})}{k^2} \label{div_E_H}
\end{eqnarray}
with phase differences  
$\phi_{\alpha\beta}(\textbf{k},t)=\phi_\beta(\textbf{k},t)-\phi_\alpha(\textbf{k},t)$,
$\alpha$ and $\beta$ standing for the component indices. Note that when one component of the vector 
${\bf k}$ is equal to zero, this case is treated separately in the code.
Since only four of these equations are independent (because of the incompressibility condition), 
we are led to give an arbitrary value 
for two of the variables.
However, the choice of these arbitrary values is constrained. Indeed, from the set of equations Eqs. (\ref{div_E_H}), 
we can derive an existence condition on the $\rho_\alpha$-amplitudes depending on the realisability condition 
($|\mathcal H({\bf k})|\le k \mathcal E({\bf k})$) and which reads (with ${\bf k}$-dependency omitted):
\begin{equation}
\big(1-\frac{k_\alpha^2}{k^2}\big)\mathcal{E}\big(1-\Gamma) \le \rho_\alpha^2 
\le \big(1-\frac{k_\alpha^2}{k^2}\big)\mathcal{E}\big(1+\Gamma\big)
\end{equation}
with $\Gamma = \sqrt{1-\mathcal{H}^2/{k^2\mathcal{E}^2}}$.
Thus, the $\rho_\alpha$-amplitudes can be expressed as:
\begin{equation}
\rho_\alpha^2 = \Big[\rho_\alpha^{i^2}-(1-\frac{k_\alpha^2}{k^2})\mathcal{E}^i\Big]
\frac{\Gamma^2}
{\Gamma^{i^2}}
 +\Big[ 1-\frac{k_\alpha^2}{k^2}\Big] \mathcal{E} \ ,
\label{B3}\end{equation}
where the $i$ superscript denotes quantities based on the intermediate velocity field, which is a
solution of the modified Navier-Stokes equation
Eq.~(\ref{NSfmodel}), with eddy viscosity and helical eddy diffusivity incorporated, and with
$\Gamma^i=\sqrt{1-\mathcal{H}^{i^2}/{k^2\mathcal{E}^{i^2}}}$.
Eq~(\ref{B3}) represents a projection of $\rho_{\alpha}^i$ (computed from the intermediate
velocity field and which depends on $\mathcal{E}^i$ and $\mathcal{H}^i$) to obtain the amplitude 
$\rho_{\alpha}({\bf k},t)$ 
(depending now on $\mathcal{E}$ and $\mathcal{H}$) at the updated time step. 
This allows not to modify the velocity field
when 
$\widehat{T}_E^{pq}=0 $ and $\widehat{T}_H^{pq}=0$. 

If one or two componants of the ${\bf k}$-wavevector are equal to zero, the set
of equations (\ref{div_E_H}) is rewritten from the divergence-free condition.  
Apart from this, the reconstruction procedure is similar.

Finally, to rebuild the different velocity phases, $\phi_{1}$ is assumed to be fixed to its value 
given by the intermediate component $v_1({\bf k},t)$. 
The set of equations (\ref{div_E_H}) is then solved and we obtain the updated Fourier velocity field.
Note that a different choice for the fixed phase leads to no significant changes in
our numerical tests.

\end{document}